\documentclass[twocolumn]{aastex63}
\usepackage{tikz}
\usetikzlibrary{shapes,arrows,chains,fit,calc,shapes,matrix,positioning}
\usepackage{multirow}
\newcommand{\angstrom}{\mbox{\normalfont\AA}}
\usepackage{siunitx}
\usepackage{subfigure}
\usepackage{pgfplots}
\pgfplotsset{compat=newest}

\submitjournal{ApJS}
\shorttitle{ISM Ions}
\shortauthors{Tinacci et al.}
\graphicspath{{./}{figures/}}
\usepackage{comment}
\usepackage{hyperref}

\begin{document}
\title{Structures and Properties of Known and Postulated Interstellar Cations}
\correspondingauthor{Lorenzo Tinacci}
\email{Lorenzo.Tinacci@univ-grenoble-alpes.fr}
\author[0000-0001-9909-9570]{Lorenzo Tinacci}
\affiliation{Dipartimento di Chimica and Nanostructured Interfaces and Surfaces (NIS) Centre,\\ Università degli Studi di Torino, via P. Giuria 7, 10125 Torino, Italy}
\affiliation{Université Grenoble Alpes, CNRS, IPAG, 38000 Grenoble, France}
\author[0000-0002-2457-1065]{Stefano Pantaleone}
\affiliation{Dipartimento di Chimica and Nanostructured Interfaces and Surfaces (NIS) Centre,\\ Università degli Studi di Torino, via P. Giuria 7, 10125 Torino, Italy}
\affiliation{Dipartimento di Chimica, Biologia e Biotecnologie, Università di Perugia, 06123 Perugia, Italy}
\author[0000-0002-5524-8068]{Andrea Maranzana}
\affiliation{Dipartimento di Chimica and Nanostructured Interfaces and Surfaces (NIS) Centre,\\ Università degli Studi di Torino, via P. Giuria 7, 10125 Torino, Italy}
\author[0000-0001-5121-5683]{Nadia Balucani}
\affiliation{Université Grenoble Alpes, CNRS, IPAG, 38000 Grenoble, France}
\affiliation{Dipartimento di Chimica, Biologia e Biotecnologie, Università di Perugia, 06123 Perugia, Italy}
\affiliation{Osservatorio Astrofisico di Arcetri, Largo E. Fermi 5, 50125 Firenze, Italy}
\author[0000-0001-9664-6292]{Cecilia Ceccarelli}
\affiliation{Université Grenoble Alpes, CNRS, IPAG, 38000 Grenoble, France}
\author[0000-0001-8886-9832]{Piero Ugliengo}
\affiliation{Dipartimento di Chimica and Nanostructured Interfaces and Surfaces (NIS) Centre,\\ Università degli Studi di Torino, via P. Giuria 7, 10125 Torino, Italy}

\begin{abstract}
Positive ions play a fundamental role in the interstellar chemistry, especially in cold environments where chemistry is believed to be mainly ion-driven.   
However, in contrast with neutral species, most of the cations present in the astrochemical reaction networks are not fully characterized in the astrochemical literature.
To fill up this gap, we have carried out new accurate quantum chemical calculations to identify the structures and energies of 262 cations with up to 14 atoms that are postulated to have a role in the interstellar chemistry. 
Optimised structures and rotational constants were obtained at M06-2X/cc-pVTZ level, while electric dipoles and total electronic energies were computed with CCSD(T)/aug-cc-pVTZ//M06-2X/cc-pVTZ single point energy calculations. 
The present work complements the study by \cite{woon2009quantum}, who characterised the structure and energies of 200 neutral species involved as well in the interstellar chemistry.
Taken together, the two datasets can be used to estimate whether a reaction, postulated in present astrochemical reaction networks, is feasible from a thermochemistry point of view and, consequently, to improve the reliability of the present networks used to simulate the interstellar chemistry.
We provide an actual example of the potential use of the cations plus neutral datasets. 
It shows that two reactions, involving Si-bearing ions and present in the widely used reaction networks KIDA and UMIST, cannot occur in cold ISM because endothermic.
\end{abstract}


\section{Introduction} \label{sec:intro}

Soon after the first detection in the late 1960s of polyatomic molecules in interstellar cold (10--20 K) molecular clouds \citep{cheung1968detection,cheung1969detection,snyder1969microwave}, it became clear the dominant role of cations in the chemistry leading to them \citep{watson1973rate,herbst1973formation}.
The reason is relatively simple: molecular clouds are too cold for reactions which present activation barriers to take place and the vast majority of neutral-neutral reactions possess activation barriers (insurmountable at 10 K).
Therefore, in cold molecular clouds, chemistry is believed to be mainly driven by cations, whose root is the ionisation of Hydrogen (both in the atomic and molecular forms) by the cosmic-rays that permeate the Milky Way.

The first confirmation of this theoretical prediction came with the detection of HCO$^+$ by \cite{snyder1976interpretation}\footnote{Actually, the theory followed the suggestion by \cite{klemperer1970carrier} that an unidentified line observed in a few sources was to be attributed to HCO$^+$.}.
To date, out of a bit more than 200 interstellar detected species, about 30 are cations  (\url{https://cdms.astro.uni-koeln.de/classic/molecules}), the last ones  discovered being HC$_3$S$^+$ and CH$_3$CO$^+$ \citep{cernicharo2021space,cernicharo2021discovery}.
Interesting, all the 13 detected cations with more than 3 atoms are, so far, protonated forms of stable and abundant molecules.
It is important to emphasise that the relatively low number of detected cations is not due to their real paucity, at least based on the astrochemical theoretical predictions, but on their low abundance and the difficulty of deriving their spectroscopic properties \citep[e.g.][]{2020NatRPcation}.

As a matter of fact, of the about 500 species involved in the present astrochemical gas-phase reaction networks (e.g. KIDA\footnote{\url{http://kida.astrophy.u-bordeaux.fr/}} and UMIST\footnote{\url{http://udfa.ajmarkwick.net/}}) more than half are cations.
On the same vein, of the about 8000 reactions in the same reaction networks the vast majority, about 5500, involves cations.
Yet, despite the obvious importance of cations in the modeling of the interstellar chemistry, no systematic study exists in the literature on the structure and energy of the cations involved in these networks.
Indeed, it is worth noticing that the above mentioned reaction networks list cations whose structure has seldomly been characterised and often appear as chemical formulae guessed on the basis of the reactions that involve them.
On the contrary, a systematic theoretical study of many neutral species present in the astrochemical reaction networks has been carried out more than a decade ago by \cite{woon2009quantum}.

The goal of the present work is to provide accurate physico-chemical data for cation species, comparable in terms of methodology with those available for neutral species, to ultimately improve the accuracy of the astrochemical models.
An obvious example of the impact of having reliable data of all species present in the astrochemical networks is that this will allow to quickly verify the exo/endo-thermicity of the reaction, and, if relevant, exclude it from the network without the need of the very time-consuming characterization of the transition states of the reaction. 

To reach the goal of providing reliable data on the cations, accurate estimates of the electronic spin multiplicity, geometrical structure and absolute electronic energy of each cation are needed \citep{herzberg1966electronic,lattelais2009interstellar,lattelais2010new,chabot2013reactions}.
Here we present new computations of the physico-chemical properties of the 262  cations present in the KIDA astrochemical gas-phase reaction network \citep{wakelam2012kinetic}. 

The article is organised as follows.
In Sect. \ref{sec:methodology}, we provide details on the adopted computational methodology.
In Sect. \ref{sec:results}, we report the results of the new computations.
In Sect. \ref{sec:discussion}, we provide two examples of the possible application of the two datasets (the neutral one from \cite{woon2009quantum} and cationic one from the present work) to identify and consequently exclude endothermic reactions present in the KIDA and UMIST reaction networks.
Sect. \ref{sec:conclusions} concludes the article including the hyperlink to the on-line database from this work which publicly available.

\section{Methodology} \label{sec:methodology}

\subsection{Initial guessed geometrical structures}\label{subsec:geom_init}
In the astrochemical reaction networks, cations fall into two general classes, whether or not they are produced by ionization of a mother neutral species. Therefore, we adopted two different approaches to recover the initial guessed structures of the 262 cations.

For the 128 cations belonging to the first class (\textit{i.e.} from ionization of a mother species), we started from the structure of the neutral species calculated by \cite{woon2009quantum} removing one electron and then optimizing the structure after assigning the proper charge and spin multiplicity. 
The reason behind this choice is that, although cations are not produced by electron abstraction processes in most cases, astrochemical networks usually postulate cations structures having the same connectivity of their neutral counterparts.

For about two third of the remaining 134 cations, we retrieved the starting structures from the KIDA database$^2$ and the NIST Computational Chemistry Comparison and Benchmark DataBase (CCCBD)\footnote{\url{http://cccbdb.nist.gov/}}.
Finally, when only the brute formulas were available (about 50 cations), we guessed the starting structure case-by-case looking at the products of the reactions forming and destroying the cation.
To automatise the initial geometric guess for the unknown chemical structure, we developed a graph theory-based software tool coupled with the UFF (Universal Force Field) \citep{rappe1992uff} implemented in RdKit \citep{landrum2016rdkit}. The script can be found in:  \href{https://aco-itn.oapd.inaf.it/aco-public-datasets/theoretical-chemistry-calculations/software-packages/cations-structures-scripts}{ACO-Cations-Scripts}.

We emphasise that the procedures and choices described above stem from the fact that, very often, only a simple connectivity or a cation name are available in the reaction networks and not the structure itself. 
In the few uncertain cases where \textit{cis} - \textit{trans} isomers are possible and no further information is available in the network to differentiate them, we assumed the most stable one (in the specific case, the \textit{trans} isomer, based on the general rules of organic chemistry).

\subsection{Computational details}
Once the 262 guessed geometrical structures were obtained, a sequence of geometric optimizations at DFT (Density Functional Theory) level considering both electronic spin multiplicities in the ground and first excited states. 

All calculations were carried out with the Gaussian16 program\citep{g16}. 
For the DFT calculations, we adopted the Minnesota method M06-2X \citep{M06-2X} coupled with the triple-$\zeta$ Dunning's correlation consistent basis set (cc-pVTZ) \citep{kendall1992dunning,woon1993gaussian} for  geometry optimization. 
We kept the default values set up in Gaussian16 for: DFT integration grid (\textit{i.e.} 99590 grid points), SCF (\textit{i.e.} $\Delta \mathrm{E} = 10^{-8}$ Ha on RMS density matrix and $\Delta \mathrm{E} = 10^{-6}$ Ha on MAX density matrix and total energy), and geometry optimization tolerances (\textit{i.e.} $3\cdot10^{-4}$ Ha and $1.2\cdot10^{-3}$ a$_0$, on RMS gradients and displacements, respectively).
We carefully explored symmetry constraints to maximize the number of symmetry elements compatible with the most stable structure. 
Harmonic frequency calculations were performed for all considered cases to ensure that each structure was a minimum of the PES (Potential Energy Surface). 
Dipole moments and absolute electronic energies were refined at coupled cluster level with full single and double excitations and perturbative treatment of triple excitations (CCSD(T) and ROCCSD(T) \citep{knowles1993coupled,watts1993coupled}) in conjunction with an augmented triple-$\zeta$ correlation consistent basis set (aug-cc-pVTZ) \citep{kendall1992dunning}.

\pgfdeclarelayer{marx}
\pgfsetlayers{main,marx}
\providecommand{\cmark}[2][]{%
  \begin{pgfonlayer}{marx}
    \node [nmark] at (c#2#1) {#2};
  \end{pgfonlayer}{marx}
  } 
\providecommand{\cmark}[2][]{\relax}
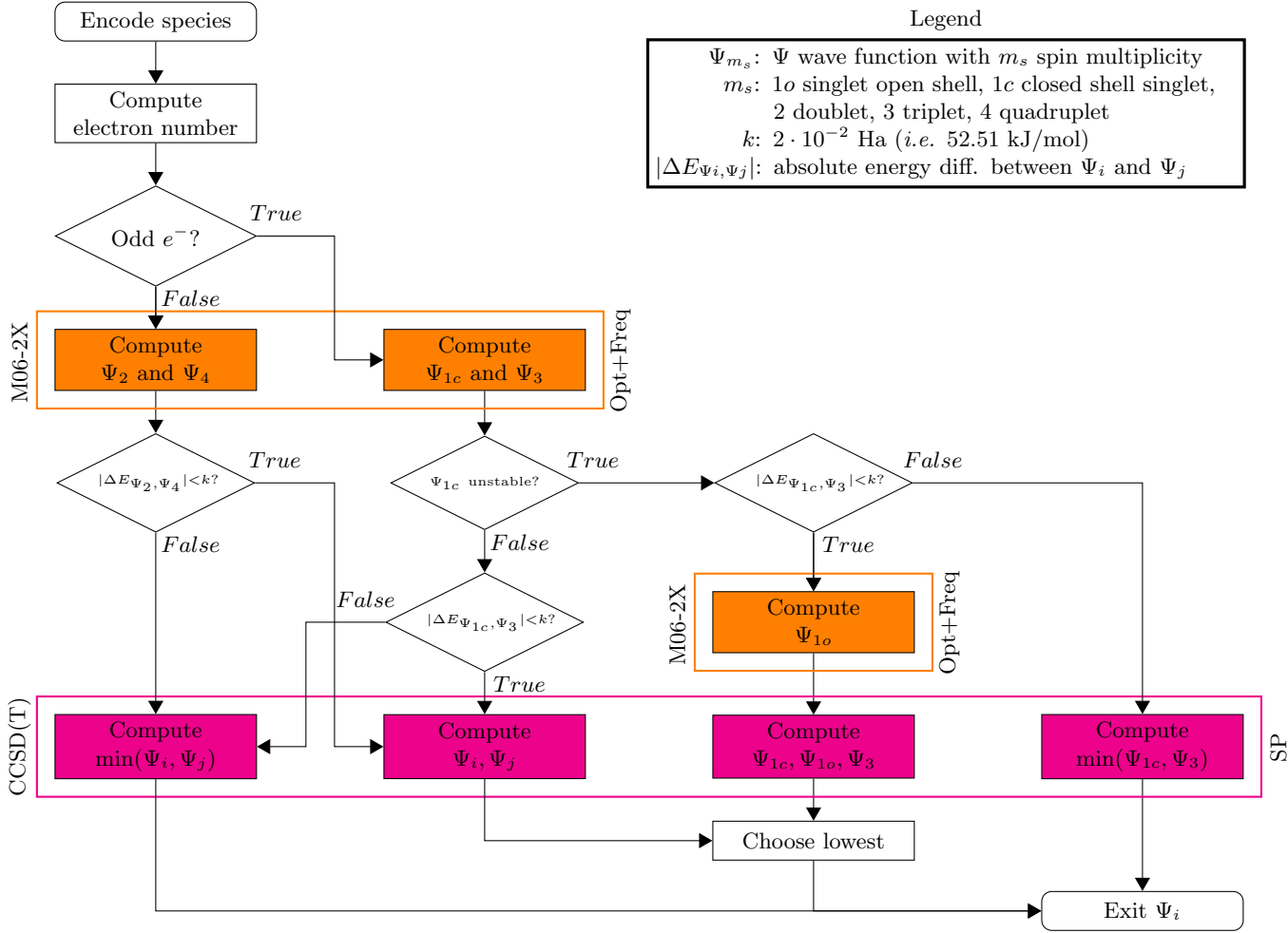
\begin{figure*}
    \centering
\begin{tikzpicture}[
    >=triangle 60,              
    start chain=going below,    
    node distance=5.9mm and 46mm, 
    every join/.style={norm},   
]
\tikzset{
  base/.style={draw, on chain, on grid, align=center, minimum height=4ex},
  proc/.style={base, rectangle, text width=8em},
  leg/.style={base, text width=25em, align=left, very thick},
  test/.style={base, diamond, aspect=2, text width=5em},
  term/.style={proc, rounded corners},
  coord/.style={coordinate, on chain, on grid, node distance=6mm and 25mm},
  nmark/.style={draw, cyan, circle, font={\sffamily\bfseries}},
  norm/.style={->, draw},
  free/.style={->, draw},
  cong/.style={->, draw},
  it/.style={font={\small\itshape}}
}
\node [term, join] (0) {Encode species};
\node [proc, join] (t0) {Compute electron number};
\node [test, join] (t1) {Odd $e^-$?};
\node [proc, fill=orange, join] (t2) {Compute $\Psi_2$ and $\Psi_4$};
\node [test, join] (t3) {\tiny{$|\Delta E_{\tiny\Psi_2,\Psi_4}|$<$k$?}};
\node [coord, right=of t1] (c1)  {};
\node [coord, right=of t3] (c2)  {};
\node [proc, fill=orange, right=of t2] (p0) {Compute $\Psi_{1c}$ and $\Psi_3$};
\node [test, join, right= of t3] (p1) {\tiny {$\Psi_{1c}$ unstable?}};
\node [test] (p2) {\tiny{$|\Delta E_{\tiny\Psi_{1c},\Psi_3}|$<$k$?}};
\node [proc, fill=magenta] (p3) {Compute\\ $\Psi_{i},\Psi_j$};
\node [proc, fill=magenta,left=of p3] (t4) {Compute $\min(\Psi_{i},\Psi_j)$};
\node [coord, left=of p2] (c3)  {};
\node [test, right=of p1] (d1) {\tiny{$|\Delta E_{\tiny \Psi_{1c},\Psi_3}|$<$k$?}};
\node [proc, fill=orange, right=of p2, join] (d0) {Compute \\$\Psi_{1o}$};
\node [proc, fill=magenta, right=of p3,join] (d2) {Compute \\$\Psi_{1c},\Psi_{1o},\Psi_3$};
\node [proc,join] (p4) {Choose lowest};
\node [coord, right=of d1] (c5)  {};
\node [coord, left=of d1] (c4)  {};
\node [proc, fill=magenta, right=of d2] (d3) {Compute $\min(\Psi_{1c},\Psi_3)$};
\node [term, join, yshift=-3em] (ex)     {Exit $\Psi_i$};
\node [left=1.5em of t2, rotate=90,xshift=2.0em] {M06-2X};
\node [left=1.5em of d0, rotate=90,xshift=2.0em] {M06-2X};
\node [right=1.5em of p0, rotate=90,xshift=-2.6em] {Opt+Freq};
\node [right=1.5em of d0, rotate=90,xshift=-2.6em] {Opt+Freq};
\node [left=1.5em of t4, rotate=90,xshift=2.5em] {CCSD(T)};
\node [right=1.5em of d3, rotate=90,xshift=-1em] {SP};
\node[draw=magenta,inner sep=7pt,thick,fit=(t4)(p3)(d2)(d3)] {};
\node[draw=orange,inner sep=7pt,thick,fit=(t2)(p0)] {};
\node[draw=orange,inner sep=7pt,thick,fit=(d0)] {};
\path (t1.east) to node [near start, yshift=1em] {$True$} (c1); 
  \draw [->] (t1.east) -- (c1) |- (p0);
\path (t1.south) to node [near start, xshift=1.5em, yshift=-0.5em] {$False$} (t1);
  \draw [->] (t1.south) -- (t2);
 \path (t3.south) to node [near start, xshift=1.5em, yshift=-0.5em] {$False$} (t3);
  \draw [->] (t3.south) -- (t4);
 \path (p1.south) to node [near start, xshift=1.5em, yshift=-0.5em] {$False$} (p1);
  \draw [->] (p1.south) -- (p2);
\path (p2.south) to node [near start, xshift=1.5em, yshift=-0.5em] {$True$} (p2);
  \draw [->] (p2.south) -- (p3);
\draw [->] (p4.south) |- (ex);
\path (t3.east) to node [near start, yshift=1em] {$True$} (c2); 
  \draw [->] (t3.east) -- (c2) |- (p3);
\path (p2.west) to node [near start, yshift=1em] {$False$} (c3);
  \draw [<-] (t4.east) -| (c3) |- (p2.west);
\path (p1.east) to node [near start, yshift=1em] {$True$} (c4); 
  \draw [->] (p1.east) -- (c4) -- (d1);
\path (d1.south) to node [near start, xshift=1.5em, yshift=-0.5em] {$True$} (d1);
  \draw [->] (d1.south) -- (d0);
\path (d1.east) to node [near start, yshift=1em] {$False$} (c5); 
  \draw [->] (d1.east) -- (c5) -| (d3.north);
\draw [->] (t4.south) |- (ex.west);
\draw [->] (p3.south) |- (p4.west);
\node [leg, right=34em of t0] (leg) {{\hspace{21pt}$\Psi_{m_s}$: $\Psi$ wave function with $m_s$ spin multiplicity\\
\hspace{27pt}$m_s$: $1o$ singlet open shell, $1c$ closed shell singlet,\\
\hspace{47pt}$2$ doublet, $3$ triplet, $4$ quadruplet\\
\hspace{34.5pt}$k$: $2\cdot 10^{-2}$ Ha (\textit{i.e.} 52.51 kJ/mol)\\
$|\Delta E_{\tiny \Psi i,\Psi j}|$: absolute energy diff. between $\Psi_i$ and $\Psi_j$}\\
};
\node [above=1.5em of leg, yshift=-1.5em] {Legend};
\end{tikzpicture}
\caption{Adopted procedure to define the ground electronic state for a species and to achieve the related optimized geometry and cation properties. 
To this end, we used two levels of theory: M06-2X/cc-pVTZ and CCSD(T)/aug-cc-pVTZ. 
In the M06-2X block, a geometric optimization (Opt) and vibrational frequencies calculation (Freq)  are performed to ensure that a minimum PES is achieved. 
In the CCSD(T) block, the final energy is then refined as a single point (SP) evaluation at CCSD(T) level together with the corresponding properties. 
The $\Psi_{1c}$ stability is tested with the Gaussian16 wave function stability tool to find Restricted $\rightarrow$ Unrestricted wave function instability.}
\label{fig:flowchart_ms}
\end{figure*}

Since unrestricted electronic solutions are affected by spin contamination (i.e. the artificial mixing of different spin states) and this contribution is not negligible nor automatically corrected in Gaussian16 via spin-annihilation procedure, we adopted the Restricted Open (RO) (except for singlet open, since the RO formalism is not applicable) formalism, whose wave function is eigenfunction of the $\mathrm{\hat{S}}^2$ operator, for all open shell configurations.
Spin contamination causes problems to recover dynamic correlation mainly in post-HF methods that are based on many-body perturbation theory (MP2, CCSD), because the perturbation through high-spin states is too large to be correctly accounted for by these methods \citep{watts1993coupled}.

Moreover, a stability analysis on the converged wave function was applied to the singlet states computed via M06-2X/cc-pVTZ level \citep{bauernschmitt1996stability} using the specific keywords (\textit{opt=stable}) provided in Gaussian16 (see Fig. \ref{fig:flowchart_ms} $\Psi_{1c}$ stability block).

The scheme summarising the adopted strategy is shown in Fig. \ref{fig:flowchart_ms}, where the various steps needed to  reach the final minimum structure are shown.

Rendering of molecule images have been obtained via the VMD software\citep{vmd} meanwhile the graphics elaboration and plots via TikZ and PGFPlots \LaTeX $\,$ packages.

\subsection{Benchmark method} \label{sec:benchmark}
In order to test the accuracy of the above described methodology, a benchmark on both structure and wave function optimizations using different methods was carried out. 

\paragraph{Geometry optimization}
We checked the M06-2X/cc-pVTZ level of theory by comparing our results obtained for a subset of molecules with those computed at CCSD/aug-cc-pVTZ level, as shown in Table \ref{tab:bench_geom}.
Both the geometry root means square deviation (RMSD) and the energy difference are small enough (less than ${\sim}0.045$ \angstrom ~ and ${\sim}1.7$ kJ/mol, respectively) to validate the present methodology. Morover the average $\Delta\mathrm{E}$ is 0.522 kJ/mol, which is lower than the commonly accepted QM calculation accuracy of ${\sim}4$ kJ/mol (\textit{i.e.} 1 kcal/mol). 

\begin{deluxetable}{lccc}
\tablehead{Species & State & $\Delta \mathrm{E}$ [kJ/mol] & RMSD [\angstrom]}
\tablecaption{Root-mean-square displacement (RMSD) of atomic positions, absolute energy difference calculated at CCSD(T)/aug-cc-pVTZ//M06-2X/cc-pVTZ compared with calculations at CCSD(T)/aug-cc-pVTZ//CCSD/aug-cc-pVTZ.}
\label{tab:bench_geom}
\startdata
$\mathrm{C}_2^+$        & $^4\Sigma_g$      & 0.247   & 0.003 \\
$\mathrm{NH}_4^+$       & $^1\mathrm{A}_1$  & -0.022  & 0.000 \\
$\mathrm{H_2CO}^+$      & $^2\mathrm{B}_2$  & 0.247   & 0.007 \\
$\mathrm{PNH}_2^+$      & $^2\mathrm{B}_2$  & 0.512   & 0.005 \\
l-$\mathrm{C_3H}_2^+$   & $^2\mathrm{A}$'   & 0.055   & 0.045 \\
$\mathrm{H_3CS}^+$      & $^3\mathrm{A}_1$  & 0.509   & 0.012 \\
c-$\mathrm{C_3H}_3^+$   & $^1\mathrm{A}_1$' & 0.714   & 0.004 \\
$\mathrm{C_4H}_3^+$     & $^1\mathrm{A}_1$  & 1.681   & 0.008 \\
$\mathrm{CH_3CHOH}^+$   & $^1\mathrm{A}$'   & 0.432   & 0.006 \\
$\mathrm{H_2C_3O}^+$    & $^2\mathrm{B}_2$  & 0.845   & 0.005 \\
\enddata
\end{deluxetable}

\paragraph{Wave function optimization}
The procedure shown in Figure \ref{fig:flowchart_ms} was first tested on two very common molecules: ethylene (C$_2$H$_6$) and methylene (CH$_2$). 
These molecules present different ground states, \textit{i.e.} singlet closed-shell ($^1\mathrm{A}_g$) and triplet ($^3\mathrm{B}_1$) for ethylene and methylene, respectively.
Their corresponding excited state are triplet ($^3\mathrm{A}_1$) for ethylene and singlet closed-shell ($^1\mathrm{A}_1$) for methylene. 
Our calculated transition energies are in good agreement with the experimental data: 282 versus 272 kJ/mol \citep{ethylene_ref} and 40 versus 38 kJ/mol \citep{methylene_ref}) for ethylene and methylene, respectively. 
For the singlet closed-shell methylene ($^1\mathrm{A}_1$) the stability analysis was also performed revealing a preference for the unrestriced solution with respect to the restricted one, as expected.

\paragraph{Dipole moment evaluation}
\cite{woon2009quantum} showed that the CCSD(T)/aug-cc-pVTZ electronic dipole for neutral species is in good agreement with the experimental data. 
We expect a similar or better agreement for cations, in virtue of the more contracted nature of the electron density compared to the more diffuse one in neutral species.

\subsection{Neutral versus cation structure connectivity}\label{subsec:graph}
Since many cation structures are derived from the neutral counterpart, we have calculated the ionization energy and followed the connectivity change (if any) after the geometry relaxation on the proper PES.
In order to check if a cation retains the same connectivity of the neutral counterpart after geometric relaxation, the graph theory approach was used.
Finally, the adiabatic ionization energy for the species that has a neutral counterpart will be presented in the dedicated subsection \ref{subsec:IE} with a comparison with available experimental data.

\begin{figure}
\begin{center}
\includegraphics[width=1.\linewidth]{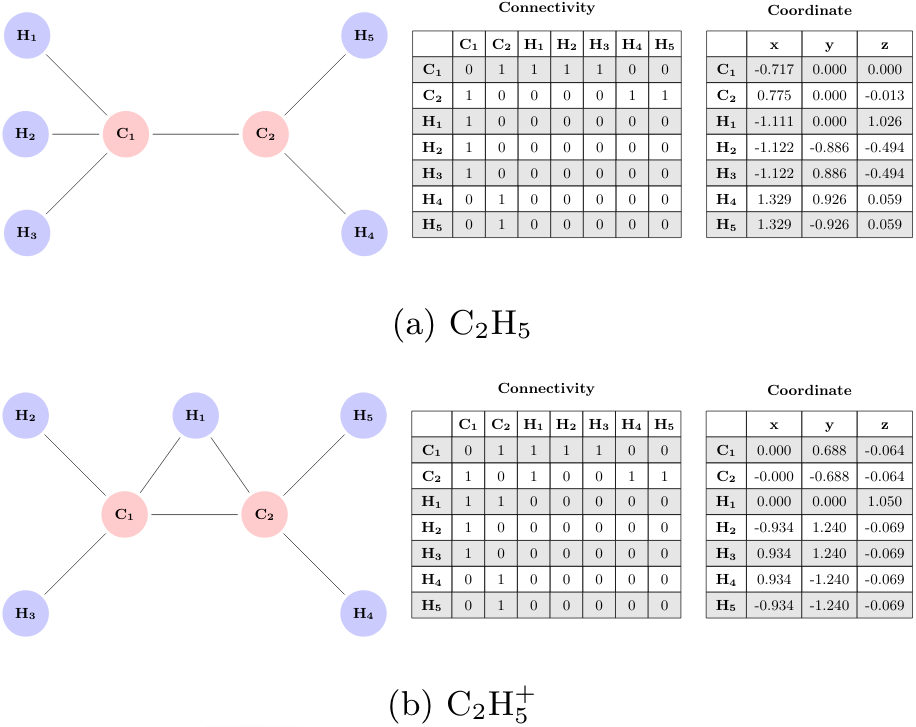}
\end{center}
\caption{Graph, connectivity matrix and coordinates (in \angstrom) of neutral ethyl radical $\mathrm{C_2H_5}$ and its cation $\mathrm{C_2H}^+_5$.}
\label{fig:graph}
\end{figure}

Firstly, all the coordinate files are converted into chemical graphs \citep{trinajstic2018chemical} using the covalent radii and distance functions implemented in the Atomic Simulation Environment (ASE) python package \citep{larsen2017atomic}. 
A chemical graph is a non-directed graph where atoms and bonds in molecules correspond to nodes and edges, respectively.
The different chemical elements present in the periodic table are represented in graphs as colors assigned to the vertices: the graph is, therefore, defined as a multi-colored graph. Secondly, the graphs related to the neutral and the ionized molecule are tested for
isomorphism. 
Two graphs which contain the same number of graph colored-vertices connected in the same way are considered isomorphic.
The NetworkX python package \citep{SciPyProceedings_11} was used in order to deal with graph objects. 
In Figure \ref{fig:graph} examples of chemical graphs are shown.

\section{Results} \label{sec:results}
The structures of the cations computed following the procedures and methodology described in the previous section are minima of PES when starting from the guessed ones, \textit{i.e.} we did not explore the full PES in search of the global minimum.
Indeed, our goal is to consider cations whose connectivity derives from the astrochemical reaction networks with structures based on the reactions giving rise to their formation.
When more isomeric forms exist, we adopted the most stable one, when specific information in the reaction network was missing.
Finally, we cross-checked our computed cation structures with the literature ones, in the relatively few cases were they are available, and generally found a very good agreement.

\subsection{Cations properties and geometries}
\begin{figure*}
\begin{center}
\subfigure[C$_2$H$_3$CO$^+$]{\includegraphics[width=0.24\linewidth]{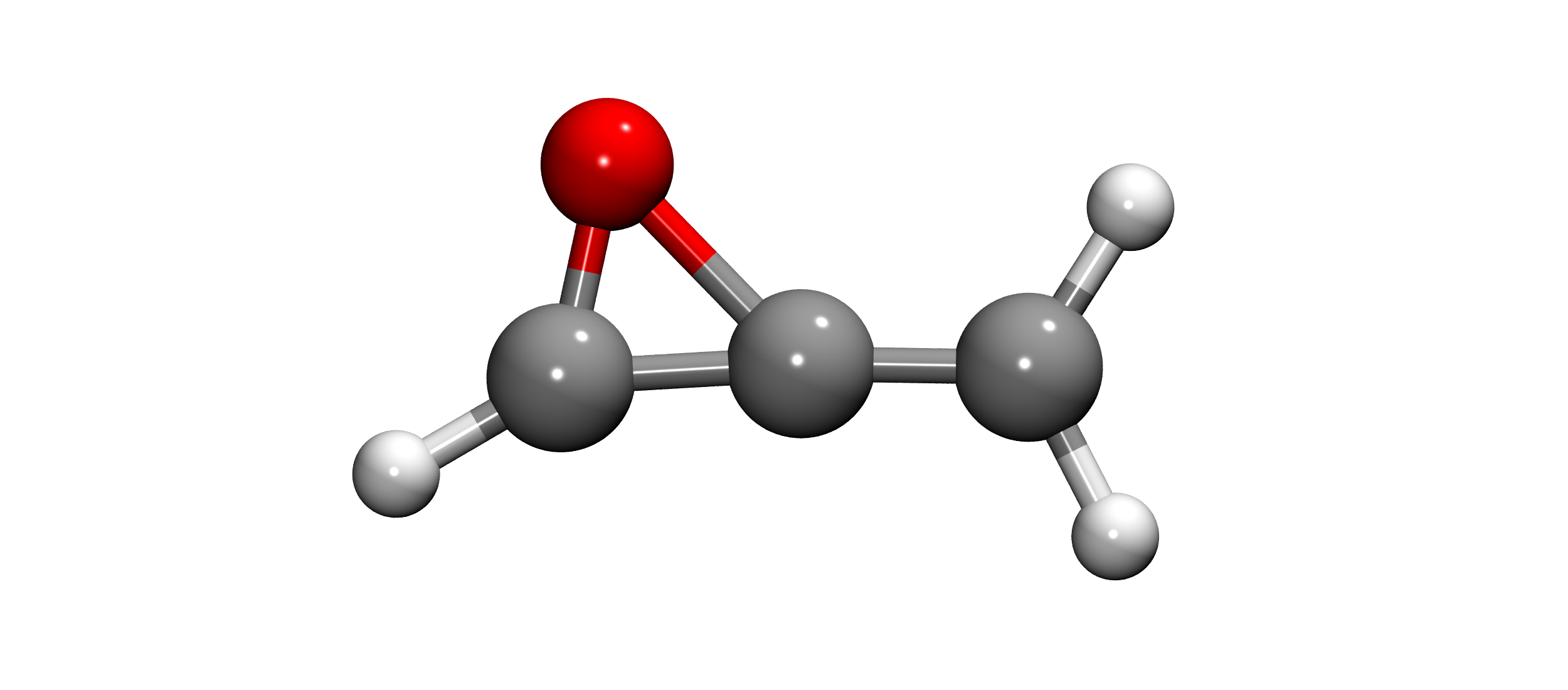}}
\subfigure[C$_2$H$_4$O$^+$]{\includegraphics[width=0.24\linewidth]{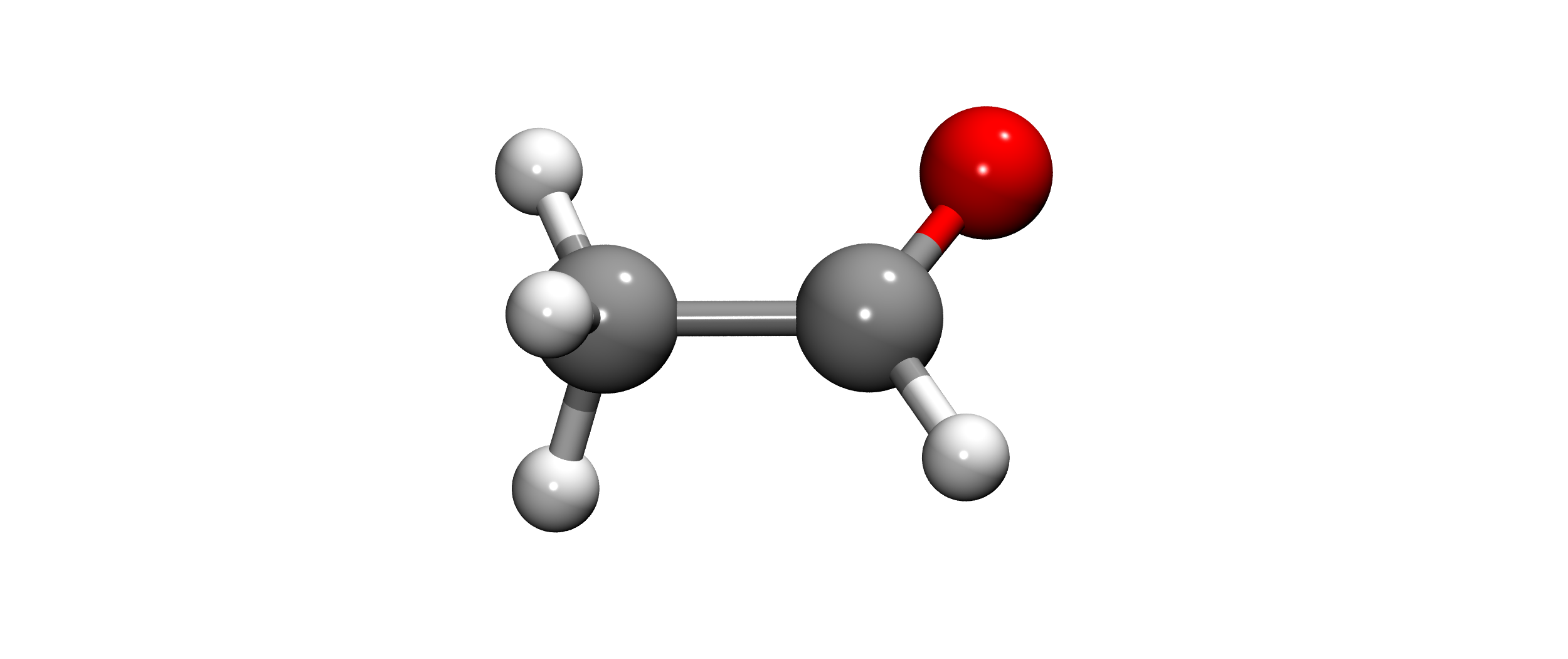}}
\subfigure[C$_3$H$_3$N$^+$]{\includegraphics[width=0.24\linewidth]{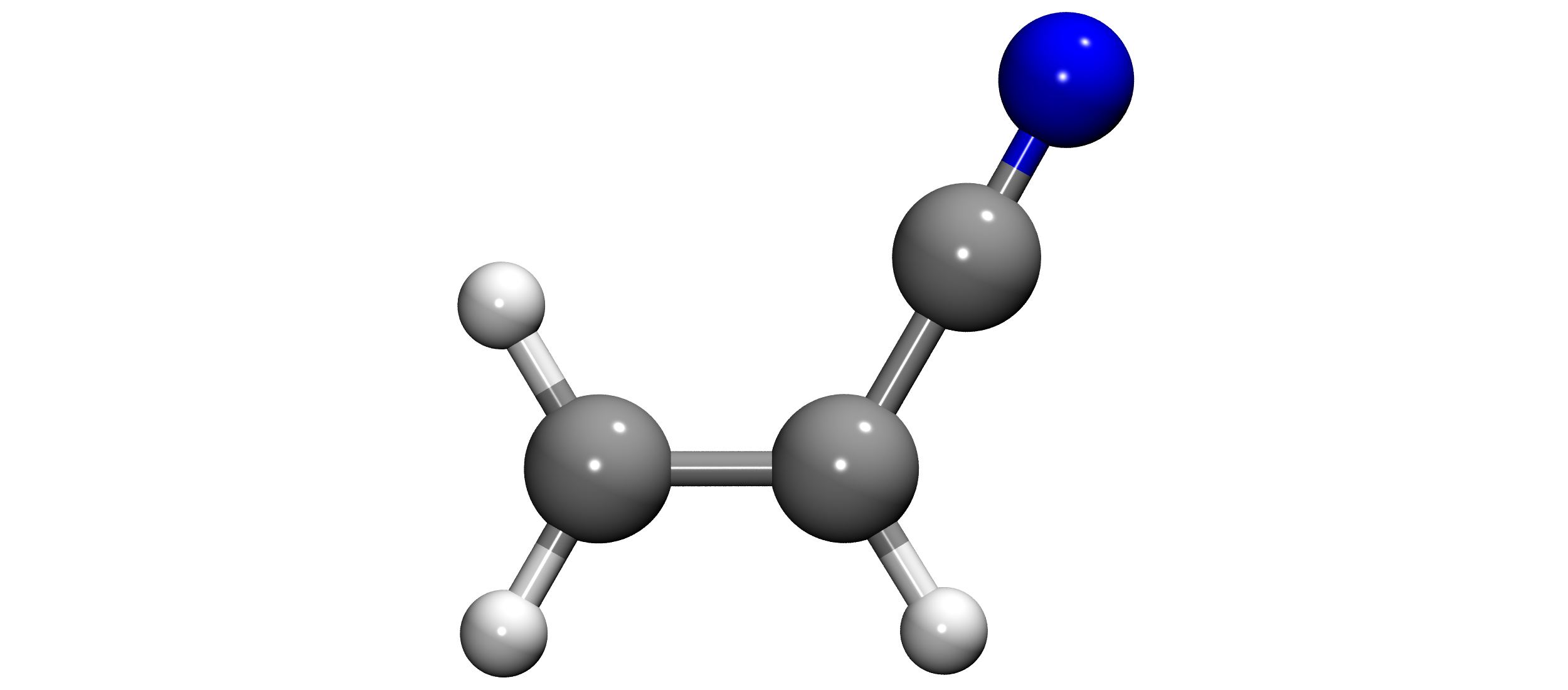}} \subfigure[C$_4$H$_3^+$]{\includegraphics[width=0.24\linewidth]{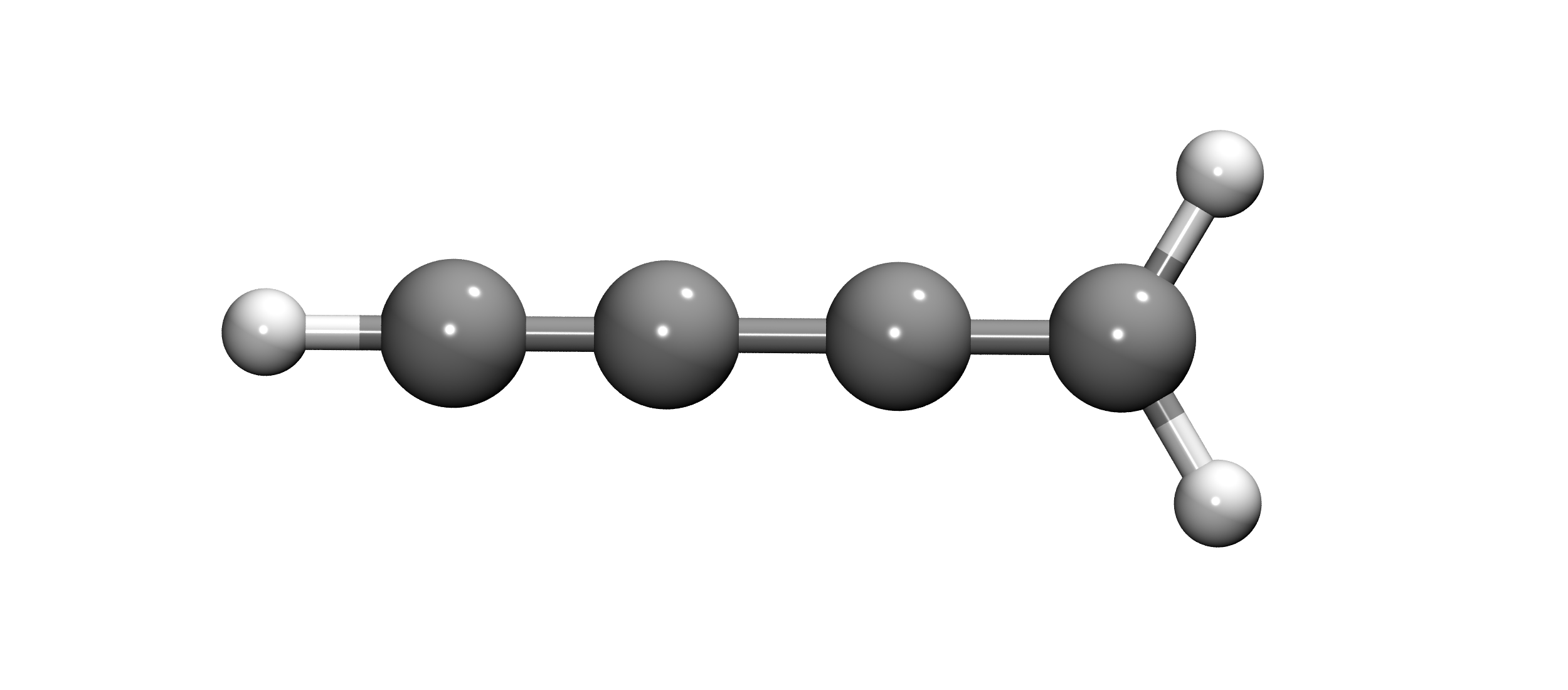}}\\
\subfigure[C$_6$H$_4^+$]{\includegraphics[width=0.24\linewidth]{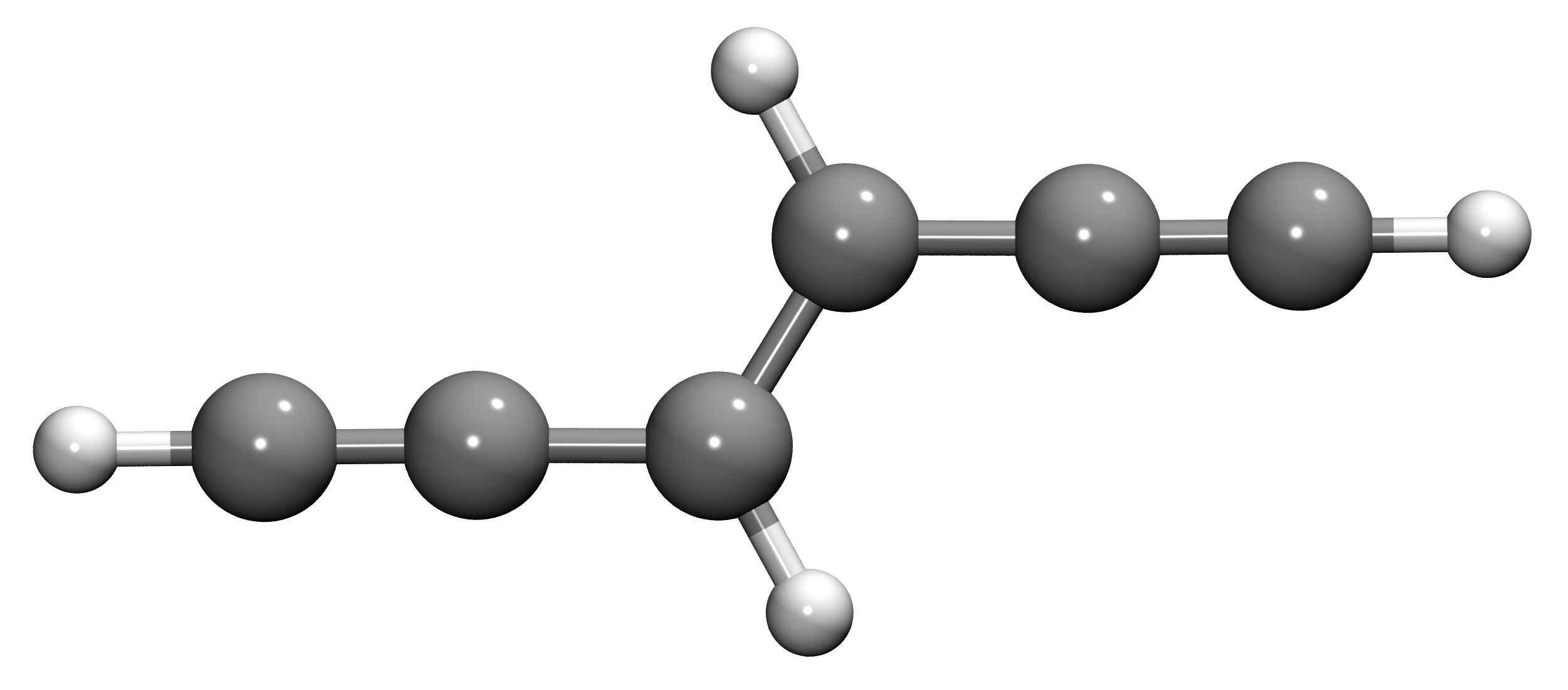}}
\subfigure[H$_3$C$_4$NH$^+$]{\includegraphics[width=0.24\linewidth]{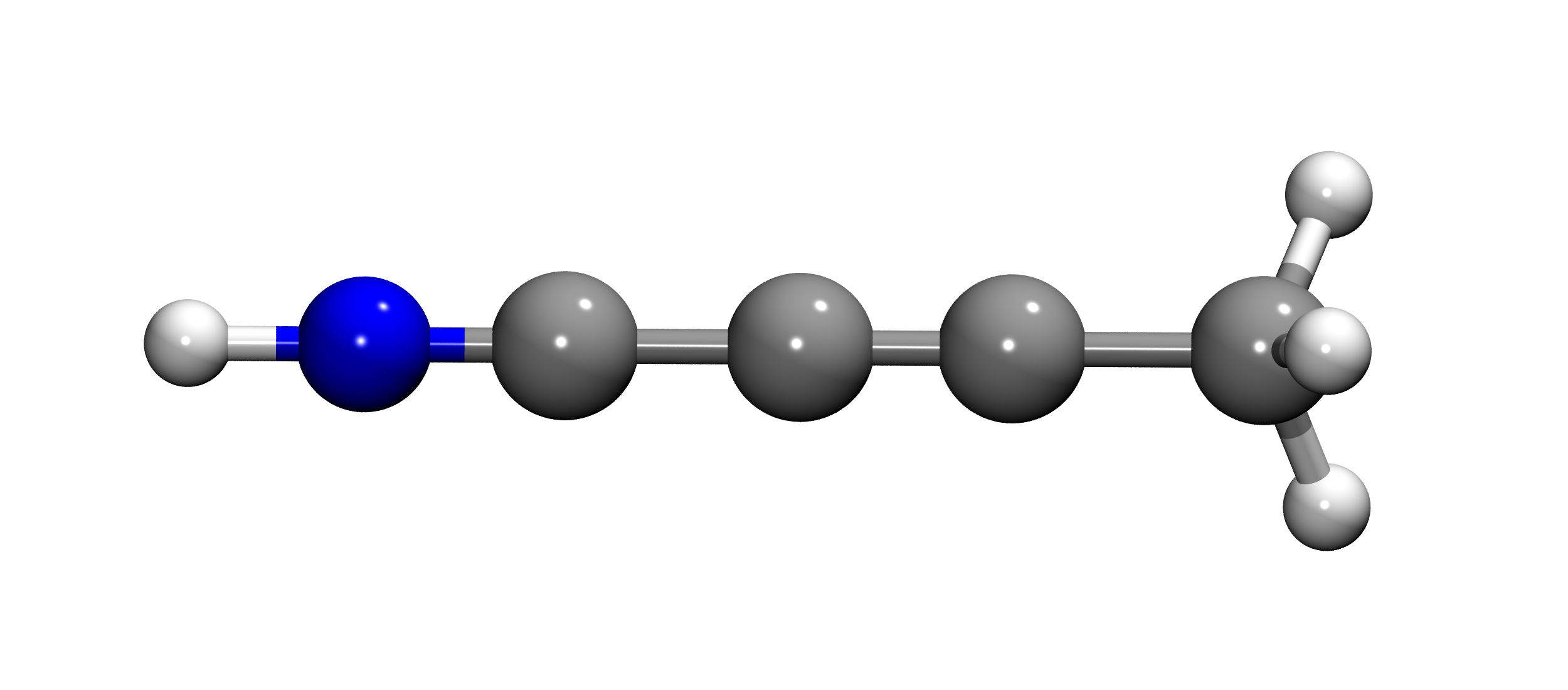}}
\subfigure[l-C$_3$H$_3^+$]{\includegraphics[width=0.24\linewidth]{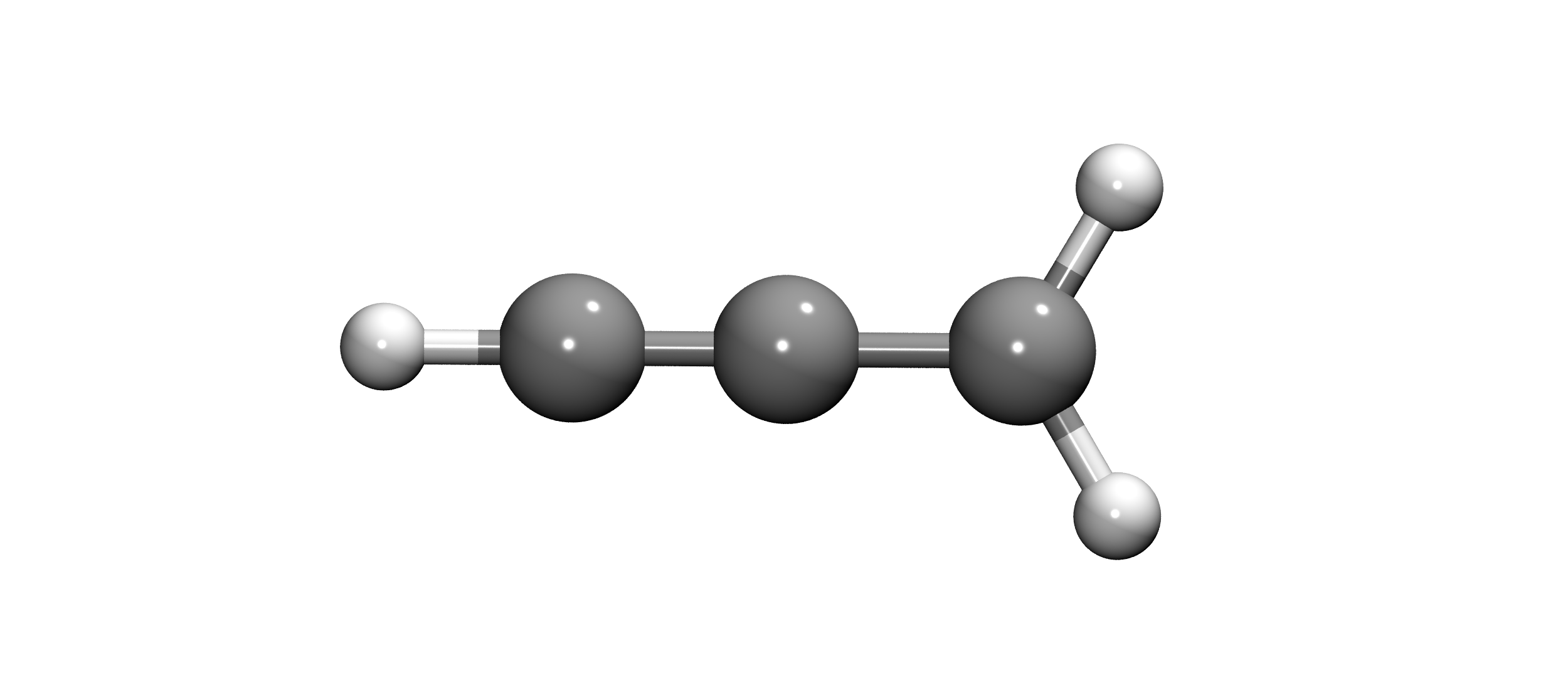}}
\subfigure[PC$_2$H$_3^+$]{\includegraphics[width=0.24\linewidth]{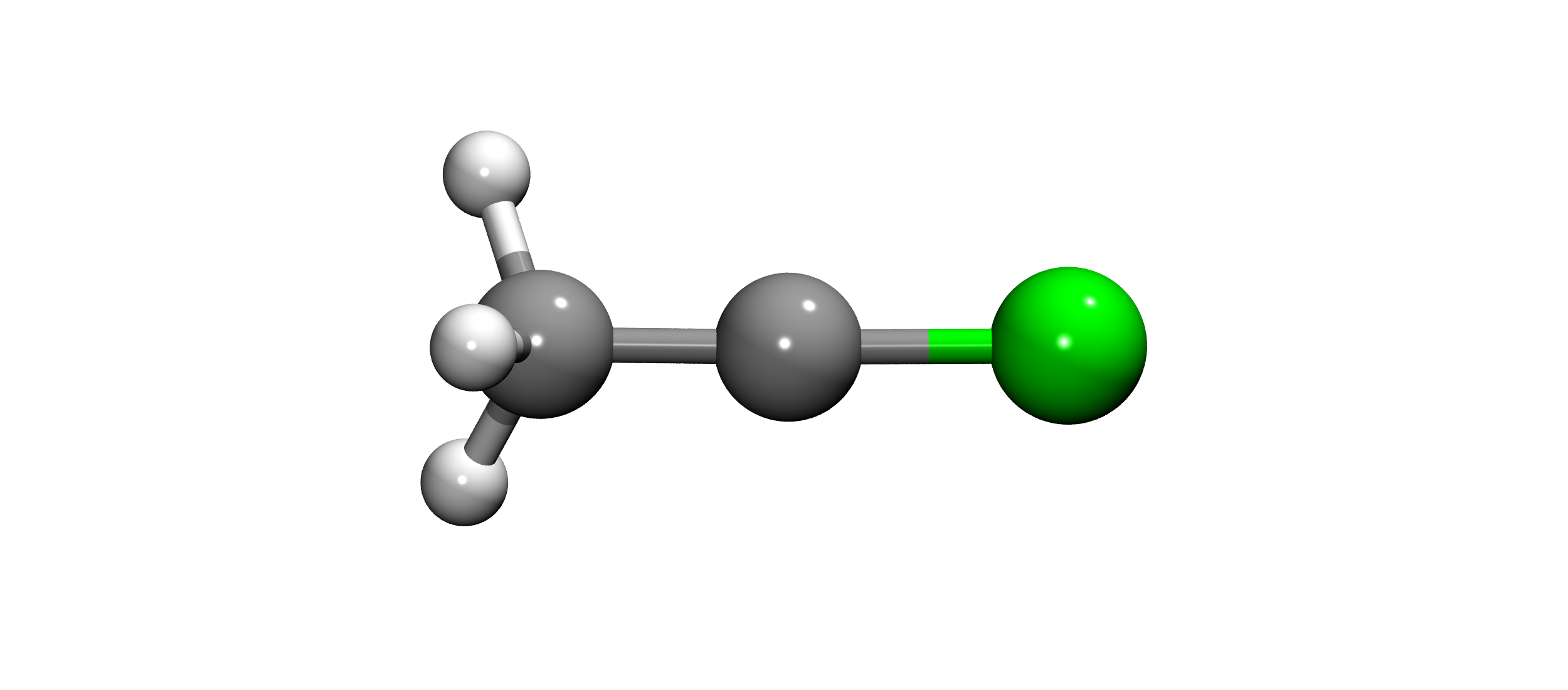}}
\end{center}
\label{fig:struct_example}
\caption{M06-2X/cc-pVTZ optimized structures of a sample of the cations studied in the present work. The structure of all the 262 cations is available in the website \href{https://aco-itn.oapd.inaf.it/aco-public-datasets/theoretical-chemistry-calculations/cations-database}{ACO-Cations-Database}}.
\end{figure*}

Tables \ref{tab:diatomic}–\ref{tab:poly} list the calculated properties of the 262 cations studied in this work. 
In the Tables, we grouped the cations into five categories: (1) diatomic species (Table \ref{tab:diatomic}); (2) linear species with three to twelve atoms (Table \ref{tab:linear}); (3) $\mathrm{C}_{2v}$ symmetry species (Table \ref{tab:c2v}); (4) planar species (excluding those belonging to point 3) (Table \ref{tab:planar}); (5) nonplanar species  (Table \ref{tab:poly}). 
Within each table, cations are ordered by increasing number of atoms and, within each subset, by increasing molecular mass. 
A sample of the derived structures is shown in Fig. \ref{fig:struct_example}.

An electronic version with extended information on all 262 cations can be found in the \textit{data.tar} file provided in the on line Supporting Material. The center of the coordinate frame with respect the dipole moment components are referred to is the centre of nuclear charge.
We also make publicly available it on website of the ACO (Astro-Chemistry Origin) project site\footnote{\url{https://aco-itn.oapd.inaf.it/home}} at the link: \href{https://aco-itn.oapd.inaf.it/aco-public-datasets/theoretical-chemistry-calculations/cations-database}{ACO-Cations-Database}. 
The web-based ACO cation structure database is based on the molecule hyperactive JSmol\footnote{JSmol is an open-source Java viewer for chemical structures in 3D: \url{http://www.jmol.org/})} plugin. 
The ACO cation structure website will be periodically updated when new species will be added to the database.
In the same ACO website, we also provide the python-script tool to convert molecules in chemical graph and to control if two species share the same connectivity by the isomorphic function.
The script can be found in \href{https://aco-itn.oapd.inaf.it/aco-public-datasets/theoretical-chemistry-calculations/software-packages/cations-structures-scripts}{ACO-Cations-Scripts}.

Finally, while the rotational constants are reported in Tables \ref{tab:diatomic}--\ref{tab:linear} and in the web-based ACO cations database (for all the other non-linear species), we warn the reader that the level of theory is not sufficiently accurate to allow their use for assigning bands in experimental spectra.

\startlongtable
\begin{deluxetable*}{lcccccc}
\tablehead{Species & State        & Symmetry         & $\mu$ & B$_e$ & $r_e$ & $r_e$(exp)}
\tablecaption{Predicted Equilibrium Structures and Properties of Diatomic Species.}
\label{tab:diatomic}
\startdata
$\mathrm{H}_2^+$    & $^2\Sigma_g$   & $\mathrm{D_{\infty h}}$ & 0.000   & 827.202  & 1.101 & 1.057$^a$ \\ 
$\mathrm{HeH}^+$    & $^1\Sigma$     & $\mathrm{C_{\infty v}}$ & 1.343   & 1010.927 & 0.788 &           \\ 
$\mathrm{CH}^+$     & $^1\Sigma$     & $\mathrm{C_{\infty v}}$ & 1.565   & 428.701  & 1.126 & 1.131$^b$ \\ 
$\mathrm{NH}^+$     & $^2\Pi$        & $\mathrm{C_{\infty v}}$ & 1.834   & 466.167  & 1.074 &           \\ 
$\mathrm{OH}^+$     & $^3\Sigma$     & $\mathrm{C_{\infty v}}$ & 2.127   & 501.498  & 1.031 & 1.029$^a$ \\ 
$\mathrm{HF}^+$     & $^2\Pi$        & $\mathrm{C_{\infty v}}$ & 2.478   & 518.136  & 1.010 &           \\ 
$\mathrm{C}_2^+$    & $^4\Sigma_g$   & $\mathrm{D_{\infty h}}$ & 0.000   & 43.077   & 1.398 &           \\ 
$\mathrm{CN}^+$     & $^1\Sigma$     & $\mathrm{C_{\infty v}}$ & 2.846   & 57.192   & 1.169 & 1.290$^a$ \\ 
$\mathrm{CO}^+$     & $^2\Sigma$     & $\mathrm{C_{\infty v}}$ & 3.286   & 60.629   & 1.103 &           \\ 
$\mathrm{N}_2^+$    & $^2\Sigma_g$   & $\mathrm{D_{\infty h}}$ & 0.000   & 59.719   & 1.099 & 1.113$^a$ \\ 
$\mathrm{SiH}^+$    & $^1\Sigma$     & $\mathrm{C_{\infty v}}$ & 0.330   & 228.988  & 1.506 & 1.499$^a$ \\ 
$\mathrm{NO}^+$     & $^1\Sigma$     & $\mathrm{C_{\infty v}}$ & 0.603   & 61.361   & 1.050 & 1.062$^a$ \\ 
$\mathrm{CF}^+$     & $^1\Sigma$     & $\mathrm{C_{\infty v}}$ & 1.305   & 51.829   & 1.151 & 1.263$^a$ \\ 
$\mathrm{PH}^+$     & $^2\Pi$        & $\mathrm{C_{\infty v}}$ & 0.741   & 255.260  & 1.424 &           \\ 
$\mathrm{O}_2^+$    & $^2\Pi_g$      & $\mathrm{D_{\infty h}}$ & 0.000   & 52.927   & 1.093 & 1.116$^a$ \\ 
$\mathrm{HS}^+$     & $^3\Sigma$     & $\mathrm{C_{\infty v}}$ & 1.170   & 278.193  & 1.364 &           \\
$\mathrm{HCl}^+$    & $^2\Pi$        & $\mathrm{C_{\infty v}}$ & 1.634   & 295.614  & 1.321 &           \\
$\mathrm{SiC}^+$    & $^4\Sigma$     & $\mathrm{C_{\infty v}}$ & 1.026   & 18.276   & 1.815 &           \\
$\mathrm{SiN}^+$    & $^3\Pi$        & $\mathrm{C_{\infty v}}$ & 2.860   & 17.154   & 1.777 &           \\
$\mathrm{CP}^+$     & $^3\Sigma$     & $\mathrm{C_{\infty v}}$ & 0.120   & 22.542   & 1.610 &           \\
$\mathrm{CS}^+$     & $^2\Sigma$     & $\mathrm{C_{\infty v}}$ & 0.945   & 26.449   & 1.480 &           \\
$\mathrm{PN}^+$     & $^2\Sigma$     & $\mathrm{C_{\infty v}}$ & 2.040   & 24.210   & 1.471 &           \\
$\mathrm{NS}^+$     & $^1\Sigma$     & $\mathrm{C_{\infty v}}$ & 2.209   & 25.735   & 1.420 &           \\
$\mathrm{PO}^+$     & $^1\Sigma$     & $\mathrm{C_{\infty v}}$ & 4.030   & 23.802   & 1.419 &           \\
$\mathrm{SiF}^+$    & $^1\Sigma$     & $\mathrm{C_{\infty v}}$ & 3.599   & 18.802   & 1.541 &           \\
$\mathrm{CCl}^+$    & $^1\Sigma$     & $\mathrm{C_{\infty v}}$ & 0.165   & 24.050   & 1.534 &           \\
$\mathrm{SiO}^+$    & $^2\Sigma$     & $\mathrm{C_{\infty v}}$ & 3.253   & 21.372   & 1.524 &           \\
$\mathrm{SO}^+$     & $^2\Pi$        & $\mathrm{C_{\infty v}}$ & 2.785   & 23.644   & 1.416 &           \\
$\mathrm{ClO}^+$    & $^3\Sigma$     & $\mathrm{C_{\infty v}}$ & 0.569   & 21.249   & 1.472 &           \\
$\mathrm{SiS}^+$    & $^2\Sigma$     & $\mathrm{C_{\infty v}}$ & 4.576   & 9.240    & 1.915 &           \\
$\mathrm{S}_2^+$    & $^2\Pi_g$      & $\mathrm{D_{\infty h}}$ & 0.000   & 9.641    & 1.811 &           \\
\enddata
\tablecomments{$\mu$ is the electric dipole moment  in Debye units. B$_e$ is the Rotational Constants express in GHz referred to the equilibrium structure. $r_e$ is the calculated internuclear equilibrium distance expressed in Angstroms, meanwhile $r_e$(exp) are the available experimental data: $^a$ from \citep{chase1996nist}; $^b$ from \citep{huber2013molecular}.}
\end{deluxetable*}

\startlongtable
\begin{deluxetable*}{lcccc|lcccc}
\tablehead{Species & State        & Symmetry         & $\mu$ & B$_e$ & Species & State        & Symmetry         & $\mu$ & B$_e$}
\tablecaption{Predicted Equilibrium Properties of Linear Polyatomic Species.}
\label{tab:linear}
\startdata
$\mathrm{C_2H}^+$   & $^3\Pi$            & $\mathrm{C_{\infty v}}$ & 0.755         & 41.665        & $\mathrm{SiC}_4^+$     & $^2\Sigma$            & $\mathrm{C_{\infty v}}$ & 7.0130  & 1.516  \\
$\mathrm{HCN}^+$    & $^2\Pi$            & $\mathrm{C_{\infty v}}$ & 3.586         & 41.202        & $\mathrm{C_4P}^+$      & $^1\Sigma$            & $\mathrm{C_{\infty v}}$ & 2.101   & 1.530  \\
$\mathrm{HNC}^+$    & $^2\Sigma$         & $\mathrm{C_{\infty v}}$ & 0.148         & 47.950        & $\mathrm{C_4S}^+$      & $^2\Pi$               & $\mathrm{C_{\infty v}}$ & 2.294   & 1.531  \\
$\mathrm{HCO}^+$    & $^1\Sigma$         & $\mathrm{C_{\infty v}}$ & 4.172         & 45.359        & $\mathrm{C_4H}_2^+$    & $^2\Pi_g$             & $\mathrm{D_{\infty h}}$ & 0.000   & 4.431  \\
$\mathrm{HOC}^+$    & $^1\Sigma$         & $\mathrm{C_{\infty v}}$ & 2.380         & 44.893        & $\mathrm{HC_2NCH}^+$   & $^1\Sigma$            & $\mathrm{C_{\infty v}}$ & 3.484   & 4.689  \\
$\mathrm{N_2H}^+$   & $^1\Sigma$         & $\mathrm{C_{\infty v}}$ & 3.154         & 47.597        & $\mathrm{HC_3NH}^+$    & $^1\Sigma$            & $\mathrm{C_{\infty v}}$ & 1.246   & 4.350  \\
$\mathrm{CNC}^+$    & $^1\Sigma_g$       & $\mathrm{D_{\infty h}}$ & 0.000         & 13.661        & $\mathrm{C_5H}^+$      & $^1\Sigma$            & $\mathrm{C_{\infty v}}$ & 2.677   & 2.420  \\
$\mathrm{C_2N}^+$   & $^1\Sigma$         & $\mathrm{C_{\infty v}}$ & 2.660         & 12.004        & $\mathrm{HC_4N}^+$     & $^2\Pi$               & $\mathrm{C_{\infty v}}$ & 6.436   & 2.327  \\
$\mathrm{CHSi}^+$   & $^3\Sigma$         & $\mathrm{C_{\infty v}}$ & 0.246         & 15.767        & $\mathrm{HC_4O}^+$     & $^3\Sigma$            & $\mathrm{C_{\infty v}}$ & 4.504   & 2.246  \\
$\mathrm{NCO}^+$    & $^3\Sigma$         & $\mathrm{C_{\infty v}}$ & 1.117         & 11.196        & $\mathrm{C}_6^+$       & $^2\Pi_u$             & $\mathrm{D_{\infty h}}$ & 0.000   & 1.445  \\
$\mathrm{HNSi}^+$   & $^2\Sigma$         & $\mathrm{C_{\infty v}}$ & 3.821         & 16.886        & $\mathrm{C_5N}^+$      & $^3\Sigma$            & $\mathrm{C_{\infty v}}$ & 3.890& 1.391\\
$\mathrm{HCP}^+$    & $^2\Pi$            & $\mathrm{C_{\infty v}}$ & 0.985         & 18.914        & $\mathrm{SiC_4H}^+$    & $^1\Sigma$            & $\mathrm{C_{\infty v}}$ & 2.487   & 1.440  \\
$\mathrm{CO}_2^+$   & $^2\Pi_g$          & $\mathrm{D_{\infty h}}$ & 0.000         & 11.600        & $\mathrm{PC_4H}^+$     & $^2\Sigma$            & $\mathrm{C_{\infty v}}$ & 0.274   & 1.444  \\
$\mathrm{HCS}^+$    & $^1\Sigma$         & $\mathrm{C_{\infty v}}$ & 1.964         & 21.653        & $\mathrm{HC_4S}^+$     & $^3\Sigma$            & $\mathrm{C_{\infty v}}$ & 2.736   & 1.463  \\
$\mathrm{HSiO}^+$   & $^1\Sigma$         & $\mathrm{C_{\infty v}}$ & 6.769         & 19.171        & $\mathrm{C_6H}^+$      & $^3\Sigma$            & $\mathrm{C_{\infty v}}$ & 3.410   & 1.393  \\
$\mathrm{HPN}^+$    & $^1\Sigma$         & $\mathrm{C_{\infty v}}$ & 0.267         & 20.749        & $\mathrm{HC_5N}^+$     & $^2\Pi$               & $\mathrm{C_{\infty v}}$ & 7.395   & 1.345  \\
$\mathrm{NO}_2^+$   & $^1\Sigma_g$       & $\mathrm{D_{\infty h}}$ & 0.000         & 12.903        & $\mathrm{HC_5O}^+$     & $^1\Sigma$            & $\mathrm{C_{\infty v}}$ & 2.586   & 1.308  \\
$\mathrm{SiC}_2^+$  & $^2\Sigma$         & $\mathrm{C_{\infty v}}$ & 1.472         & 6.125         & $\mathrm{C}_7^+$       & $^2\Sigma_u$          & $\mathrm{D_{\infty h}}$ & 0.000   & 0.912  \\
$\mathrm{SiNC}^+$   & $^1\Sigma$         & $\mathrm{C_{\infty v}}$ & 4.015         & 6.653         & $\mathrm{C_6H}_2^+$    & $^2\Pi_u$             & $\mathrm{D_{\infty h}}$ & 0.000   & 1.343  \\
$\mathrm{C_2S}^+$   & $^2\Pi$            & $\mathrm{C_{\infty v}}$ & 1.112         & 6.489         & $\mathrm{H_2C_5N}^+$   & $^1\Sigma$            & $\mathrm{C_{\infty v}}$ & 3.489   & 1.300  \\
$\mathrm{OCS}^+$    & $^2\Pi$            & $\mathrm{C_{\infty v}}$ & 1.653         & 5.849         & $\mathrm{C_7H}^+$      & $^1\Sigma$            & $\mathrm{C_{\infty v}}$ & 2.299   & 0.885  \\
$\mathrm{HSiS}^+$   & $^1\Sigma$         & $\mathrm{C_{\infty v}}$ & 5.168         & 8.661         & $\mathrm{HC_6N}^+$     & $^2\Pi$               & $\mathrm{C_{\infty v}}$ & 7.813   & 0.850  \\
$\mathrm{C_2H}_2^+$ & $^2\Pi_u$          & $\mathrm{D_{\infty h}}$ & 0.000         & 33.595        & $\mathrm{C}_8^+$       & $^2\Sigma_g$          & $\mathrm{D_{\infty h}}$ & 0.000   & 0.610  \\
$\mathrm{HCNH}^+$   & $^1\Sigma$         & $\mathrm{C_{\infty v}}$ & 0.566         & 37.543        & $\mathrm{C_7N}^+$      & $^3\Sigma$            & $\mathrm{C_{\infty v}}$ & 4.860   & 0.587  \\
$\mathrm{C_3H}^+$   & $^1\Sigma$         & $\mathrm{C_{\infty v}}$ & 2.552         & 11.262        & $\mathrm{C_8H}^+$      & $^3\Sigma$            & $\mathrm{C_{\infty v}}$ & 3.461   & 0.592  \\
$\mathrm{HSiNH}^+$  & $^1\Sigma$         & $\mathrm{C_{\infty v}}$ & 3.460         & 17.494        & $\mathrm{HC_7N}^+$     & $^2\Pi$               & $\mathrm{C_{\infty v}}$ & 8.764   & 0.570  \\
$\mathrm{C}_4^+$    & $^2\Pi_g$          & $\mathrm{D_{\infty h}}$ & 0.000         & 4.836         & $\mathrm{HC_7O}^+$     & $^1\Sigma$            & $\mathrm{C_{\infty v}}$ & 1.022   & 0.555  \\
$\mathrm{C_3O}^+$   & $^2\Sigma$         & $\mathrm{C_{\infty v}}$ & 2.809         & 4.879         & $\mathrm{C}_9^+$       & $^2\Sigma_u$          & $\mathrm{C_{\infty v}}$ & 0.011   & 0.429  \\
$\mathrm{C_2N}_2^+$ & $^2\Pi_g$          & $\mathrm{C_{\infty v}}$ & 0.000         & 4.716         & $\mathrm{C_8N}^+$      & $^1\Sigma$            & $\mathrm{C_{\infty v}}$ & 7.020   & 0.414  \\
$\mathrm{SiC_2H}^+$ & $^1\Sigma$         & $\mathrm{C_{\infty v}}$ & 0.623         & 5.582         & $\mathrm{C_8H}_2^+$    & $^2\Pi_g$             & $\mathrm{D_{\infty h}}$ & 0.000   & 0.574  \\
$\mathrm{PC_2H}^+$  & $^2\Pi$            & $\mathrm{C_{\infty v}}$ & 0.219         & 5.803         & $\mathrm{C_7H_2N}^+$   & $^1\Sigma$            & $\mathrm{C_{\infty v}}$ & 6.191   & 0.556  \\
$\mathrm{HC_2S}^+$  & $^3\Sigma$         & $\mathrm{C_{\infty v}}$ & 2.502         & 6.049         & $\mathrm{C_9H}^+$      & $^1\Sigma$            & $\mathrm{C_{\infty v}}$ & 1.405   & 0.418  \\
l-$\mathrm{SiC}_3^+$& $^2\Pi$            & $\mathrm{C_{\infty v}}$ & 1.129         & 2.651         & $\mathrm{HC_8N}^+$     & $^2\Pi$               & $\mathrm{C_{\infty v}}$ & 8.870   & 0.403  \\
$\mathrm{C_3S}^+$   & $^2\Sigma$         & $\mathrm{C_{\infty v}}$ & 0.962         & 2.858         & $\mathrm{C}_{10}^+$    & $^2\Pi_u$             & $\mathrm{D_{\infty h}}$ & 0.000   & 0.312  \\
$\mathrm{C_4H}^+$   & $^3\Sigma$         & $\mathrm{C_{\infty v}}$ & 3.140         & 4.640         & $\mathrm{C_9N}^+$      & $^3\Sigma$            & $\mathrm{C_{\infty v}}$ & 5.942   & 0.301  \\
$\mathrm{HC_3N}^+$  & $^2\Pi$            & $\mathrm{C_{\infty v}}$ & 6.378         & 4.568         & $\mathrm{C_{10}H}^+$   & $^3\Sigma$            & $\mathrm{C_{\infty v}}$ & 3.258   & 0.304  \\
$\mathrm{HC_3O}^+$  & $^1\Sigma$         & $\mathrm{C_{\infty v}}$ & 3.635         & 4.484         & $\mathrm{C9HN}^+$      & $^2\Pi$               & $\mathrm{C_{\infty v}}$ & 10.189  & 0.294  \\
$\mathrm{C}_5^+$    & $^2\Sigma_u$       & $\mathrm{D_{\infty h}}$ & 0.000         & 2.596         & $\mathrm{HC_9O}^+$     & $^1\Sigma$            & $\mathrm{C_{\infty v}}$ & 1.055   & 0.286  \\
$\mathrm{C_4N}^+$   & $^1\Sigma$         & $\mathrm{C_{\infty v}}$ & 3.900         & 2.438         & $\mathrm{C}_{11}^+$    & $^2\Sigma_u$          & $\mathrm{D_{\infty h}}$ & 0.000   & 0.235  \\
$\mathrm{SiC_3H}^+$ & $^3\Sigma$         & $\mathrm{C_{\infty v}}$ & 0.661         & 2.556         & $\mathrm{C_{10}N}^+$   & $^1\Sigma$            & $\mathrm{C_{\infty v}}$ & 8.995   & 0.227  \\
$\mathrm{PC_3H}^+$  & $^2\Pi$            & $\mathrm{C_{\infty v}}$ & 1.002         & 2.662         & $\mathrm{C_{10}H}_2^+$ & $^2\Pi_u$             & $\mathrm{D_{\infty h}}$ & 0.000   & 0.296  \\
$\mathrm{HC_3S}^+$  & $^1\Sigma$         & $\mathrm{C_{\infty v}}$ & 1.892         & 2.753         & $\mathrm{HC_{10}N}^+$  & $^2\Pi$               & $\mathrm{C_{\infty v}}$ & 9.935   & 0.222  \\
\enddata
\tablecomments{$\mu$ is the electric dipole moment  in Debye units. B$_e$ is the Rotational Constants express in GHz referred to the equilibrium structure. In the few cases in which Gaussian16 is unable to determine the electronic symmetry state for radical species exhibiting partially filled degenerate orbitals: the electronic symmetry state is labelled with an "X".}
\end{deluxetable*}

\startlongtable
\begin{deluxetable*}{lcc|lcc|lcc}
\tablehead{Species & State & $\mu$ & Species & State & $\mu$ & Species & State & $\mu$}
\tablecaption{Predicted Equilibrium Properties of Planar Polyatomic Species with $\mathrm{C}_{2v}$ Symmetry.}
\label{tab:c2v}
\startdata
$\mathrm{CH}_2^+$       & $^2\mathrm{A}_1$ & 0.488         & $\mathrm{H_2SiO}^+$     & $^2\mathrm{B}_2$ & 3.737  & $\mathrm{C_4H}_3^+$     & $^1\mathrm{A}_1$ & 0.297         \\
$\mathrm{NH}_2^+$       & $^3\mathrm{B}_1$ & 0.559         & $\mathrm{PNH}_2^+$      & $^2\mathrm{B}_2$ & 1.140  & $\mathrm{PC_4H}_2^+$    & $^1\mathrm{A}_1$ & 3.268         \\
$\mathrm{H_2O}^+$       & $^2\mathrm{B}_1$ & 2.201         & $\mathrm{H_2PO}^+$      & $^1\mathrm{A}_1$ & 5.344  & $\mathrm{C_5H}_2^+$     & $^2\mathrm{B}_2$ & 4.696         \\
$\mathrm{H_2F}^+$       & $^1\mathrm{A}_1$ & 2.342         & $\mathrm{H_2CCl}^+$     & $^1\mathrm{A}_1$ & 3.330  & $\mathrm{H_2C_4N}^+$    & $^1\mathrm{A}_1$ & 8.037         \\
$\mathrm{NaH}_2^+$      & $^1\mathrm{A}_1$ & 1.294         & $\mathrm{HSO}_2^+$      & $^1\mathrm{A}_1$ & 3.802  & $\mathrm{C_3H}_5^+$     & $^1\mathrm{A}_1$ & 0.815         \\
$\mathrm{SiH}_2^+$      & $^2\mathrm{A}_1$ & 0.116         & $\mathrm{H_2S}_2^+$     & $^2\mathrm{A}_2$ & 2.154  & $\mathrm{C_5H}_3^+$     & $^1\mathrm{A}_1$ & 1.907         \\
$\mathrm{PH}_2^+$       & $^1\mathrm{A}_1$ & 0.915         & $\mathrm{C_2H}_3^+$     & $^1\mathrm{A}_1$ & 0.785  & $\mathrm{CH_3OCH}_3^+$  & $^2\mathrm{B}_1$ & 0.791         \\
$\mathrm{H_2S}^+$       & $^2\mathrm{B}_1$ & 1.466         & c-$\mathrm{C_3H}_2^+$   & $^2\mathrm{A}_1$ & 1.128  & $\mathrm{C_6H}_3^+$     & $^1\mathrm{A}_1$ & 1.721         \\
$\mathrm{C}_3^+$        & $^2\mathrm{B}_2$ & 0.731         & $\mathrm{CH_2CN}^+$     & $^1\mathrm{A}_1$ & 5.472  & $\mathrm{C_7H}_2^+$     & $^2\mathrm{B}_2$ & 6.388         \\
$\mathrm{H_2Cl}^+$      & $^1\mathrm{A}_1$ & 2.010         & $\mathrm{H_2CCN}^+$     & $^1\mathrm{A}_1$ & 5.471  & $\mathrm{H_2C_6N}^+$    & $^1\mathrm{A}_1$ & 10.568        \\
$\mathrm{SO_2}^+$       & $^2\mathrm{A}_1$ & 1.871         & $\mathrm{H_2CCO}^+$     & $^2\mathrm{B}_1$ & 3.539  & $\mathrm{C_7H}_3^+$     & $^1\mathrm{A}_1$ & 3.426         \\
$\mathrm{CH}_4^+$       & $^2\mathrm{B}_2$ & 1.478         & $\mathrm{SiC_2H}_2^+$   & $^2\mathrm{B}_2$ & 0.038  & $\mathrm{C_6H}_5^+$     & $^1\mathrm{A}_1$ & 1.330         \\
$\mathrm{H_2NC}^+$      & $^1\mathrm{A}_1$ & 2.335         & $\mathrm{PC_2H}_2^+$    & $^1\mathrm{A}_1$ & 1.973  & $\mathrm{C_8H}_3^+$     & $^1\mathrm{A}_1$ & 3.505         \\
$\mathrm{H_2CO}^+$      & $^2\mathrm{B}_2$ & 3.025         & $\mathrm{CH_2NH}_2^+$   & $^1\mathrm{A}_1$ & 0.112  & $\mathrm{C_9H}_2^+$     & $^2\mathrm{B}_2$ & 8.177         \\
$\mathrm{H_2NO}^+$      & $^1\mathrm{A}_1$ & 3.639         & l-$\mathrm{C_3H}_3^+$   & $^1\mathrm{A}_1$ & 0.727  & $\mathrm{H_2C_8N}^+$    & $^1\mathrm{A}_1$ & 13.241        \\
$\mathrm{NaH_2O}^+$     & $^1\mathrm{A}_1$ & 2.463         & $\mathrm{NH_2CNH}^+$    & $^1\mathrm{A}_1$ & 0.180  & $\mathrm{C_9H}_3^+$     & $^1\mathrm{A}_1$ & 5.329         \\
$\mathrm{CH_2Si}^+$     & $^2\mathrm{A}_1$ & 2.280         & $\mathrm{H_2C_3O}^+$    & $^2\mathrm{B}_2$ & 4.275  & $\mathrm{C_9H_2N}^+$    & $^1\mathrm{A}_1$ & 13.766        \\
$\mathrm{PCH}_2^+$      & $^1\mathrm{A}_1$ & 0.588         & $\mathrm{SiC_3H}_2^+$   & $^2\mathrm{B}_1$ & 0.681  & $\mathrm{C_{10}H}_3^+$  & $^1\mathrm{A}_1$ & 5.697         \\
$\mathrm{H_2CS}^+$      & $^2\mathrm{B}_2$ & 2.013         & $\mathrm{C_2H}_5^+$     & $^1\mathrm{A}_1$ & 0.612  & $\mathrm{H_2C_{10}N}^+$ & $^1\mathrm{A}_1$ & 16.153        \\
\enddata   
\tablecomments{$\mu$ is the electric dipole moment  in Debye units.}
\end{deluxetable*}

\startlongtable
\begin{deluxetable*}{lccc|lccc}
\tablehead{Species & State  & Symmetry      & $\mu$ & Species & State  & Symmetry      & $\mu$}
\tablecaption{Predicted equilibrium properties of planar polyatomic species without $\mathrm{C}_{2v}$ Symmetry.}
\label{tab:planar}
\startdata
$\mathrm{H}_3^+$    & $^1\mathrm{A}_1$' & $\mathrm{D}_{3h}$ & 0.000         & $\mathrm{HSiO}_2^+$       & $^1\mathrm{A}$'   & $\mathrm{C}_s$    & 5.494   \\
$\mathrm{HNO}^+$    & $^2\mathrm{A}$'   & $\mathrm{C}_s$    & 2.908         & $\mathrm{H_2COH}^+$       & $^1\mathrm{A}$'   & $\mathrm{C}_s$    & 2.352   \\
$\mathrm{HO}_2^+$   & $^3\mathrm{A}$    & $\mathrm{C}_s$    & 2.170         & l-$\mathrm{C_3H}_2^+$     & $^2\mathrm{A}$'   & $\mathrm{C}_s$    & 3.107   \\
$\mathrm{HNS}^+$    & $^2\mathrm{A}$'   & $\mathrm{C_s}$    & 1.439         & $\mathrm{HCOOH}^+$        & $^2\mathrm{A}$'   & $\mathrm{C}_s$    & 0.432   \\
$\mathrm{C_2O}^+$   & $^2\mathrm{A}$    & $\mathrm{C}_s$    & 1.733         & $\mathrm{H_3SiO}^+$       & $^1\mathrm{A}$'   & $\mathrm{C}_s$    & 2.278   \\
$\mathrm{HPO}^+$    & $^2\mathrm{A}$'   & $\mathrm{C}_s$    & 2.962         & $\mathrm{PNH}_3^+$        & $^1\mathrm{A}$'   & $\mathrm{C}_s$    & 3.093   \\
$\mathrm{HSO}^+$    & $^1\mathrm{A}$'   & $\mathrm{C}_s$    & 3.336         & c-$\mathrm{C_3H}_3^+$     & $^1\mathrm{A}_1$' & $\mathrm{D}_{3h}$ & 0.000   \\
$\mathrm{CCP}^+$    & $^1\mathrm{A}$'   & $\mathrm{C}_s$    & 2.381         & $\mathrm{SiC_2H}_3^+$     & $^1\mathrm{A}$'   & $\mathrm{C}_s$    & 0.224   \\
$\mathrm{HS}_2^+$   & $^1\mathrm{A}$'   & $\mathrm{C}_s$    & 1.496         & $\mathrm{NH_2CH_2O}^+$    & $^1\mathrm{A}$'   & $\mathrm{C}_s$    & 2.190   \\
$\mathrm{CH}_3^+$   & $^1\mathrm{A}_1$' & $\mathrm{D}_{3h}$ & 0.000         & $\mathrm{C_3H_3N}^+$      & $^2\mathrm{A}$"   & $\mathrm{C}_s$    & 6.293   \\
$\mathrm{NH}_3^+$   & $^2\mathrm{A}_2$" & $\mathrm{D}_{3h}$ & 0.000         & c-$\mathrm{C_3H_2OH}^+$   & $^1\mathrm{A}$'   & $\mathrm{C}_s$    & 2.492   \\
$\mathrm{SiH}_3^+$  & $^1\mathrm{A}_1$' & $\mathrm{D}_{3h}$ & 0.000         & $\mathrm{HCCCHOH}^+$      & $^1\mathrm{A}$'   & $\mathrm{C}_s$    & 1.234   \\
$\mathrm{HC_2N}^+$  & $^2\mathrm{A}$'   & $\mathrm{C}_s$    & 5.002         & $\mathrm{C_2H_3CO}^+$     & $^1\mathrm{A}$'   & $\mathrm{C}_s$    & 2.743   \\
$\mathrm{C_2HO}^+$  & $^1\mathrm{A}$'   & $\mathrm{C}_s$    & 3.384         & $\mathrm{C_6N}^+$         & $^1\mathrm{A}$'   & $\mathrm{C}_s$    & 5.324   \\
$\mathrm{HNCO}^+$   & $^2\mathrm{A}$"   & $\mathrm{C}_s$    & 3.446         & $\mathrm{C_4H}_4^+$       & $^2\mathrm{A}$"   & $\mathrm{C}_s$    & 1.078   \\
$\mathrm{HOCO}^+$   & $^1\mathrm{A}$'   & $\mathrm{C}_s$    & 3.446         & $\mathrm{C_3H_3NH}^+$     & $^1\mathrm{A}$'   & $\mathrm{C}_s$    & 1.707   \\
$\mathrm{HN_2O}^+$  & $^1\mathrm{A}$'   & $\mathrm{C}_s$    & 3.706         & $\mathrm{COOCH}_4^+$      & $^2\mathrm{A}$'   & $\mathrm{C}_s$    & 3.196   \\
$\mathrm{C_3N}^+$   & $^3\mathrm{A}$"   & $\mathrm{C}_s$    & 2.921         & $\mathrm{C_5H}_5^+$       & $^1\mathrm{A}$'   & $\mathrm{C}_s$    & 1.219   \\
$\mathrm{SiNCH}^+$  & $^2\mathrm{A}$'   & $\mathrm{C}_s$    & 2.156         & $\mathrm{C_6H}_4^+$       & $^2\mathrm{A}_u$  & $\mathrm{C}_{2h}$ & 0.000   \\
$\mathrm{HOCS}^+$   & $^1\mathrm{A}$'   & $\mathrm{C}_s$    & 2.274         & & & & \\
\enddata
\tablecomments{$\mu$ is the electric dipole moment  in Debye units.}
\end{deluxetable*}

\startlongtable
\begin{deluxetable*}{lccc|lccc}
\tablehead{Species & State   & Symmetry     & $\mu$ & Species & State   & Symmetry     & $\mu$}
\tablecaption{Predicted equilibrium properties of nonplanar polyatomic species.}
\label{tab:poly}
\startdata
$\mathrm{H_3O}^+$        & $^1\mathrm{A}_1$  & $\mathrm{C}_{3v}$ & 1.414          & $\mathrm{CH_3NH}_3^+$    & $^1\mathrm{A}_1$  & $\mathrm{C}_{3v}$ & 2.166   \\
$\mathrm{PH}_3^+$        & $^2\mathrm{A}_1$  & $\mathrm{C}_{3v}$ & 0.359          & $\mathrm{CH_3CHOH}^+$    & $^1\mathrm{A}$'   & $\mathrm{C}_s$    & 2.569   \\
$\mathrm{H_3S}^+$        & $^1\mathrm{A}_1$  & $\mathrm{C}_{3v}$ & 1.554          & $\mathrm{H_3C_4N}^+$     & $^2\mathrm{A}$'   & $\mathrm{C}_s$    & 5.327   \\
$\mathrm{NH}_4^+$        & $^1\mathrm{A}_1$  & $\mathrm{T}_{d}$  & 0.000          & $\mathrm{C_2H_5OH}^+$    & $^2\mathrm{A}$    & $\mathrm{C}_{1}$  & 2.083   \\
$\mathrm{SiH}_4^+$       & $^2\mathrm{A}$'   & $\mathrm{C}_s$    & 1.222          & $\mathrm{C_4H}_5^+$      & $^1\mathrm{A}$'   & $\mathrm{C}_s$    & 0.860   \\
$\mathrm{SiCH}_3^+$      & $^1\mathrm{A}_1$  & $\mathrm{C}_{3v}$ & 0.750          & $\mathrm{H_5C_2O}_2^+$   & $^1\mathrm{A}$'   & $\mathrm{C}_s$    & 0.972   \\
$\mathrm{H_3CS}^+$       & $^3\mathrm{A}_1$  & $\mathrm{C}_{3v}$ & 0.559          & $\mathrm{C_5H}_4^+$      & $^2\mathrm{B}_3$  & $\mathrm{D}_{2}$  & 0.000   \\
$\mathrm{PCH}_3^+$       & $^2\mathrm{A}$"   & $\mathrm{C}_s$    & 0.249          & $\mathrm{H_3C_4NH}^+$    & $^1\mathrm{A}_1$  & $\mathrm{C}_{3v}$ & 2.635   \\
$\mathrm{H_3S}_2^+$      & $^1\mathrm{A}$    & $\mathrm{C}_1$    & 2.292          & $\mathrm{HCOCH_2OH}_2^+$ & $^1\mathrm{A}$'   & $\mathrm{C}_s$    & 2.474   \\
$\mathrm{CH}_5^+$        & $^1\mathrm{A}$'   & $\mathrm{C}_s$    & 1.630          & $\mathrm{C_5H_3N}^+$     & $^2\mathrm{A}$"   & $\mathrm{C}_s$    & 6.270   \\
$\mathrm{C_2H}_4^+$      & $^2\mathrm{B}_3$  & $\mathrm{D}_{2}$  & 0.000          & $\mathrm{C_2H_5OH}_2^+$  & $^1\mathrm{A}$    & $\mathrm{C}_{1}$  & 3.312   \\
$\mathrm{CH_3OH}^+$      & $^2\mathrm{A}$"   & $\mathrm{C}_s$    & 1.394          & $\mathrm{CH_3OCH}_4^+$   & $^1\mathrm{A}$'   & $\mathrm{C}_s$    & 1.177   \\
$\mathrm{CH_3CO}^+$      & $^1\mathrm{A}_1$  & $\mathrm{C}_{3v}$ & 2.977          & $\mathrm{C_2H_6CO}^+$    & $^2\mathrm{B}$    & $\mathrm{C}_{2}$  & 1.567   \\
$\mathrm{SiH}_5^+$       & $^1\mathrm{A}$'   & $\mathrm{C}_s$    & 1.284          & $\mathrm{C_5H_4N}^+$     & $^1\mathrm{A}$    & $\mathrm{C}_{1}$  & 5.401   \\
$\mathrm{SiCH}_4^+$      & $^2\mathrm{A}$'   & $\mathrm{C}_s$    & 1.224          & $\mathrm{C_4H}_7^+$      & $^1\mathrm{A}$'   & $\mathrm{C}_s$    & 1.480   \\
$\mathrm{PCH}_4^+$       & $^3\mathrm{A}$"   & $\mathrm{C}_s$    & 0.858          & $\mathrm{C_3H_6OH}^+$    & $^1\mathrm{A}$    & $\mathrm{C}_{1}$  & 1.636   \\
$\mathrm{PC_2H}_3^+$     & $^2\mathrm{A}$'   & $\mathrm{C}_s$    & 0.375          & $\mathrm{C_7H}_4^+$      & $^2\mathrm{B}_2$  & $\mathrm{D}_{2}$  & 0.000   \\
$\mathrm{CH_3O}_2^+$     & $^1\mathrm{A}$    & $\mathrm{C}_{1}$  & 4.737          & $\mathrm{H_3C_6NH}^+$    & $^1\mathrm{A}_1$  & $\mathrm{C}_{3v}$ & 4.646   \\
$\mathrm{CH_3CN}^+$      & $^2\mathrm{A}$'   & $\mathrm{C}_s$    & 2.604          & $\mathrm{H_3C_7N}^+$     & $^2\mathrm{A}$    & $\mathrm{C}_{1}$  & 5.120   \\
$\mathrm{CH_3NH}_2^+$    & $^2\mathrm{A}$'   & $\mathrm{C}_s$    & 1.974          & $\mathrm{C_7H}_5^+$      & $^1\mathrm{A}$'   & $\mathrm{C}_s$    & 2.172   \\
$\mathrm{CH_3OH}_2^+$    & $^1\mathrm{A}$'   & $\mathrm{C}_s$    & 1.835          & $\mathrm{C_8H}_4^+$      & $^2\mathrm{B}_2$  & $\mathrm{D}_{2}$  & 0.000   \\
$\mathrm{C_3H}_4^+$      & $^2\mathrm{B}_2$  & $\mathrm{D}_{2}$  & 0.000          & $\mathrm{C_6H}_7^+$      & $^1\mathrm{A}$'   & $\mathrm{C}_s$    & 0.780   \\
$\mathrm{CH_3CNH}^+$     & $^1\mathrm{A}_1$  & $\mathrm{C}_{3v}$ & 1.036          & $\mathrm{C_8H}_5^+$      & $^1\mathrm{A}$'   & $\mathrm{C}_s$    & 3.950   \\
$\mathrm{C_2H_4O}^+$     & $^2\mathrm{A}$'   & $\mathrm{C}_s$    & 2.056          & $\mathrm{C_9H}_4^+$      & $^2\mathrm{A}$    & $\mathrm{C}_{1}$  & 0.391   \\
$\mathrm{PC_2H}_4^+$     & $^1\mathrm{A}$'   & $\mathrm{C}_s$    & 0.573          & $\mathrm{C_8H_4N}^+$     & $^1\mathrm{A}_1$  & $\mathrm{C}_{3v}$ & 7.166   \\
$\mathrm{C_2H}_6^+$      & $^2\mathrm{A}_g$  & $\mathrm{C}_{2h}$ & 0.000          & $\mathrm{C_9H}_5^+$      & $^1\mathrm{A}$'   & $\mathrm{C}_{s}$  & 3.960   \\
\enddata
\tablecomments{$\mu$ is the electric dipole moment  in Debye units.}
\end{deluxetable*}

\subsection{Adiabatic Ionization Energy}\label{subsec:IE}
Table \ref{tab:ionization} shows the adiabatic ionization energies (IE) for neutral species computed at CCSD(T)/aug-cc-pVTZ by \cite{woon2009quantum} along with those of the cation counterparts. All computed IE are without ZPE (Zero Point Energy) correction.
Note that the latter are part of the group whose structures are derived from the ionization of neutral species, described in subsection \ref{subsec:geom_init}.
As few data on larger molecules provided by \cite{woon2009quantum} are computed with a lower-quality basis set, we recomputed the energy for those species with the aug-cc-pVTZ.

The Root Mean Squared Error (RMSE) between the computed ionization energies and the experimental available data is 0.308 eV.
This value validates our procedure and results since the experimental data are vertical IE meanwhile our are adiabatic IE (\textit{i.e.} the cationic species are relaxed in the proper PES).

\startlongtable
\begin{deluxetable*}{lcc|lcc|lcc|lcc}
\tablehead{Species & eV & eV$_{\mathrm{(exp)}}$ & Species & eV & eV$_{\mathrm{(exp)}}$ & Species & eV & eV$_{\mathrm{(exp)}}$ & Species  & eV & eV$_{\mathrm{(exp)}}$}
\tablecaption{Adiabatic Ionization Energy of the Corresponding Neutral Species.}
\label{tab:ionization}
\startdata
$\mathrm{H}_2^+$    & 15.528  & 15.426$^a$     & $\mathrm{HNC}^+$    & 11.961 & 12.500$^{ai}$  & $\mathrm{C_3N}^+$     & 11.816 &               & $\mathrm{C_6H}^+$      & 9.336  &               \\
$\mathrm{CH}^+$     & 10.562  & 10.640$^b$     & $\mathrm{HCO}^+$    & 8.037  & 8.140$^{al}$   & $\mathrm{C_3O}^+$     & 10.772 &               & $\mathrm{HC_5N}^+$     & 10.558 &               \\
$\mathrm{NH}^+$     & 13.420  & 13.490$^c$     & $\mathrm{SiH}_2^+$  & 9.101  &                & $\mathrm{SiC_2H}^+$   & 7.289  &               & $\mathrm{C_7}^+$       & 10.293 &               \\
$\mathrm{OH}^+$     & 12.899  & 13.017$^d$     & $\mathrm{HNO}^+$    & 10.165 & 10.100$^{am}$  & $\mathrm{C_3S}^+$     & 10.188 &               & $\mathrm{C_6N}^+$      & 8.866  &               \\
$\mathrm{HF}^+$     & 15.998  & 15.980$^e$     & $\mathrm{PH}_2^+$   & 9.826  & 9.824$^m$      & $\mathrm{CH}_4^+$     & 12.749 & 12.610$^{bh}$ & $\mathrm{C_2H}_6^+$    & 11.619 & 11.570$^{bw}$ \\
$\mathrm{C}_2^+$    & 11.728  & 11.920$^f$     & $\mathrm{HO}_2^+$   & 11.306 & 11.350$^{an}$  & $\mathrm{C_2H}_3^+$   & 8.678  & 8.250$^{bi}$  & $\mathrm{C_3H}_5^+$    & 8.029  &               \\
$\mathrm{CN}^+$     & 13.715  & 14.170$^b$     & $\mathrm{H_2S}^+$   & 10.451 & 10.453$^{ao}$  & $\mathrm{SiH}_4^+$    & 11.050 & 11.200$^{bl}$ & $\mathrm{CH_3CHOH}^+$  & 6.662  &               \\
$\mathrm{CO}^+$     & 13.959  & 14.014$^g$     & $\mathrm{C}_3^+$    & 11.606 & 11.610$^{ap}$  & c-$\mathrm{C_3H}_2^+$ & 9.118  & 9.150$^{bm}$  & $\mathrm{C_5H}_3^+$    & 8.124  &               \\
$\mathrm{N}_2^+$    & 15.533  & 15.581$^h$     & $\mathrm{C_2N}^+$   & 10.730 & 12.000$^{aq}$  & $\mathrm{CH_2CN}^+$   & 10.211 & 10.300$^{bn}$ & $\mathrm{C_6H}_2^+$    & 9.477  &               \\
$\mathrm{SiH}^+$    & 7.900   & 7.890$^i$      & $\mathrm{C_2O}^+$   & 10.912 &                & $\mathrm{H_2CCO}^+$   & 9.538  & 9.614$^{bo}$  & $\mathrm{C_7H}^+$      & 8.062  &               \\
$\mathrm{NO}^+$     & 9.191   & 9.264$^l$      & $\mathrm{NCO}^+$    & 11.620 & 11.759$^{ar}$  & $\mathrm{SiCH}_3^+$   & 7.066  &               & $\mathrm{HC_6N}^+$     & 2.494  &               \\
$\mathrm{PH}^+$     & 10.160  & 10.149$^m$     & $\mathrm{HNSi}^+$   & 11.718 &                & $\mathrm{HCOOH}^+$    & 11.282 & 11.310$^{bp}$ & $\mathrm{C}_8^+$       & 11.376 &               \\
$\mathrm{O}_2^+$    & 12.053  & 12.070$^n$     & $\mathrm{HCP}^+$    & 10.786 & 10.790$^{as}$  & $\mathrm{C_4H}^+$     & 10.057 &               & $\mathrm{C_7N}^+$      & 9.659  &               \\
$\mathrm{HS}^+$     & 10.310  & 10.421$^o$     & $\mathrm{CO}_2^+$   & 13.765 & 13.778$^{at}$  & $\mathrm{HC_3N}^+$    & 11.620 &               & $\mathrm{CH_3OCH}_3^+$ & 10.004 & 10.025$^{bz}$ \\
$\mathrm{HCl}^+$    & 12.716  & 12.790$^p$     & $\mathrm{HCS}^+$    & 7.588  & 7.412$^{au}$   & $\mathrm{C}_5^+$      & 11.116 & 12.300$^{bq}$ & $\mathrm{C_4H}_5^+$    & 7.066  &               \\
$\mathrm{SiC}^+$    & 8.853   & 9.000$^q$      & $\mathrm{NO}_2^+$   & 9.458  & 9.600$^{av}$   & $\mathrm{C_4N}^+$     & 9.514  &               & $\mathrm{C_8H}^+$      & 8.836  &               \\
$\mathrm{SiN}^+$    & 10.342  &                & $\mathrm{HPO}^+$    & 10.548 &                & $\mathrm{SiC_3H}^+$   & 8.181  &               & $\mathrm{HC_7N}^+$     & 9.825  &               \\
$\mathrm{CP}^+$     & 10.914  & 10.500$^r$     & $\mathrm{SiNC}^+$   & 7.851  &                & $\mathrm{SiC}_4^+$    & 10.099 &               & $\mathrm{C}_9^+$       & 3.734  &               \\
$\mathrm{CS}^+$     & 11.381  & 11.330$^s$     & $\mathrm{C_2S}^+$   & 10.186 &                & $\mathrm{C_4P}^+$     & 8.447  &               & $\mathrm{C_8N}^+$      & 8.430  &               \\
$\mathrm{PN}^+$     & 11.933  & 11.880$^t$     & $\mathrm{OCS}^+$    & 11.189 & 11.185$^{aw}$  & $\mathrm{C_4S}^+$     & 9.199  &               & $\mathrm{C_2H_6CO}^+$  & 9.675  & 9.700$^{ca}$  \\
$\mathrm{NS}^+$     & 8.916   & 8.870$^u$      & $\mathrm{HSiS}^+$   & 8.260  &                & $\mathrm{C_2H}_4^+$   & 10.485 & 10.510$^{br}$ & $\mathrm{C_5H}_5^+$    & 7.718  &               \\
$\mathrm{PO}^+$     & 8.476   & 8.390$^v$      & $\mathrm{SO}_2^+$   & 12.680 & 12.500$^{az}$  & $\mathrm{CH_3OH}^+$   & 10.928 & 10.850$^{bs}$ & $\mathrm{C_6H}_4^+$    & 8.993  &               \\
$\mathrm{CCl}^+$    & 8.834   & 8.900$^w$      & $\mathrm{HS}_2^+$   & 9.376  &                & $\mathrm{CH_3CN}^+$   & 12.246 & 12.201$^{bt}$ & $\mathrm{C_8H}_2^+$    & 8.978  &               \\
$\mathrm{SiO}^+$    & 11.521  & 11.300$^z$     & $\mathrm{CH}_3^+$   & 9.736  & 9.843$^{ba}$   & $\mathrm{C_4H}_2^+$   & 10.131 &               & $\mathrm{C_9H}^+$      & 7.742  &               \\
$\mathrm{SO}^+$     & 10.428  & 10.294$^{aa}$  & $\mathrm{NH}_3^+$   & 10.148 & 10.020$^{bb}$  & $\mathrm{C_5H}^+$     & 8.358  &               & $\mathrm{HC_8N}^+$     & 8.387  &               \\
$\mathrm{SiS}^+$    & 10.526  & 10.530$^{ab}$  & $\mathrm{C_2H}_2^+$ & 11.330 & 11.410$^{bc}$  & $\mathrm{HC_4N}^+$    & 9.425  &               & $\mathrm{C}_{10}^+$    & 10.647 &               \\
$\mathrm{S}_2^+$    & 9.503   & 9.400$^{ac}$   & $\mathrm{H_2CO}^+$  & 10.900 & 10.880$^{bd}$  & $\mathrm{C}_6^+$      & 12.441 &               & $\mathrm{C_9N}^+$      & 9.168  &               \\
$\mathrm{CH}_2^+$   & 10.338  & 10.350$^{ad}$  & $\mathrm{SiH}_3^+$  & 8.109  & 8.170$^{be}$   & $\mathrm{C_5N}^+$     & 10.705 &               & $\mathrm{C_{10}H}^+$   & 8.510  &               \\
$\mathrm{NH}_2^+$   & 11.131  & 10.780$^{ae}$  & $\mathrm{HNCO}^+$   & 11.557 & 11.595$^{bf}$  & $\mathrm{SiC_4H}^+$   & 7.082  &               & $\mathrm{C}_{11}^+$    & 8.984  &               \\
$\mathrm{H_2O}^+$   & 12.574  & 12.650$^{af}$  & $\mathrm{H_2CS}^+$  & 9.411  & 9.376$^{bg}$   & $\mathrm{C_2H}_5^+$   & 8.063  & 8.117$^{bu}$  & $\mathrm{C_{10}N}^+$   & 8.026  &               \\
$\mathrm{C_2H}^+$   & 11.267  & 11.610$^{ag}$  & $\mathrm{H_2SiO}^+$ & 10.674 &                & $\mathrm{CH_3NH}_2^+$ & 9.061  & 8.900$^{bv}$  & $\mathrm{C_{10}H}_2^+$ & 8.722  &               \\
$\mathrm{HCN}^+$    & 13.543  & 13.590$^{ah}$  & $\mathrm{C}_4^+$    & 10.734 &                & $\mathrm{C_4H}_3^+$   & 7.960  &               & $\mathrm{HC_{10}N}^+$  & 8.094  &               \\
\enddata
\tablecomments{Adiabatic ionization energy (eV) without ZPE correction, compared with the vertical experimental ionization energy (eV$_{\mathrm{(exp)}}$): $^a$ \citep{shiner1993h}, $^b$ \citep{huber2013molecular}, $^c$ \citep{dyke1980hei}, $^d$ \citep{wiedmann1992rotationally}, $^e$ \citep{tiedemann1979proton}, $^f$ \citep{C2H6+}, $^g$ \citep{erman1993direct}, $^h$ \citep{trickl1989state}, $^i$ \citep{boo1987reaction}, $^l$ \citep{reiser1988ionization}, $^m$ \citep{berkowitz1989photoionization}, $^n$ \citep{tonkyn1989rotationally}, $^o$ \citep{HS+}, $^p$ \citep{HCl+}, $^q$ \citep{SiC+}, $^r$ \citep{CP+}, $^s$ \citep{CS+}, $^t$ \citep{PN+}, $^u$ \citep{NS+}, $^v$ \citep{PO+}, $^w$ \citep{CCl+}, $^z$ \citep{SiO+}, $^{aa}$ \citep{SO+}, $^{ab}$ \citep{SiS+}, $^{ac}$ \citep{S2+}, $^{ad}$ \citep{CH2+},$^{ae}$ \citep{NH3+},$^{af}$ \citep{H2O+},$^{ag}$ \citep{C2H+},$^{ah}$ \citep{HCN+},$^{ai}$ \citep{HNC+},$^{al}$ \citep{HCO+},$^{am}$ \citep{HNO+},$^{an}$ \citep{HO2+},$^{ao}$ \citep{H2S+},$^{ap}$ \citep{C3+_bis} only the upperlimit value was reported,$^{aq}$ \citep{HNO+},$^{ar}$ \citep{HCO+},$^{as}$ \citep{HCP+},$^{at}$ \citep{CO2+},$^{au}$ \citep{HCS+},$^{av}$ \citep{NO2+},$^{aw}$ \citep{CO2+},$^{az}$ \citep{H2O+}, $^{ba}$ \citep{CH3+}, $^{bb}$ \citep{NH3+}, $^{bc}$ \citep{C2H2+}, $^{bd}$ \citep{H2CO+},$^{be}$ \citep{SiH3+},$^{bf}$ \citep{HNCO+},$^{bg}$ \citep{HCS+},$^{bh}$ \citep{CH4+},$^{bi}$ \citep{C2H3+},$^{bl}$ \citep{SiH4+},$^{bm}$ \citep{c-C3H2+},$^{bn}$ \citep{CH2CN+},$^{bo}$ \citep{H2CCO+},$^{bp}$ \citep{HCOOH+},$^{bq}$ \citep{C3+},$^{br}$ \citep{H2CO+},$^{bs}$ \citep{CH3OH+},$^{bt}$ \citep{CH3CN+},$^{bu}$ \citep{C2H5+},$^{bv}$ \citep{CH3NH2+},$^{bw}$ \citep{C2H6+},$^{bz}$ \citep{CH3OCH3+},$^{ca}$ \citep{C2H6CO+}. The calculated adiabatic ionization energy for C$_4$, C$_6$,  C$_8$ and  C$_{10}$ have been computed with respect to the triplet state of these species and not to the singlet state, as reported in the \cite{woon2009quantum} paper.}
\end{deluxetable*}

\begin{figure}
\centering
\begin{center}
\begin{tikzpicture}[node distance=5em, auto]
\node [fill=white] (e0) {};
\node [left=2em of e0] (net) {\includegraphics[scale=0.05]{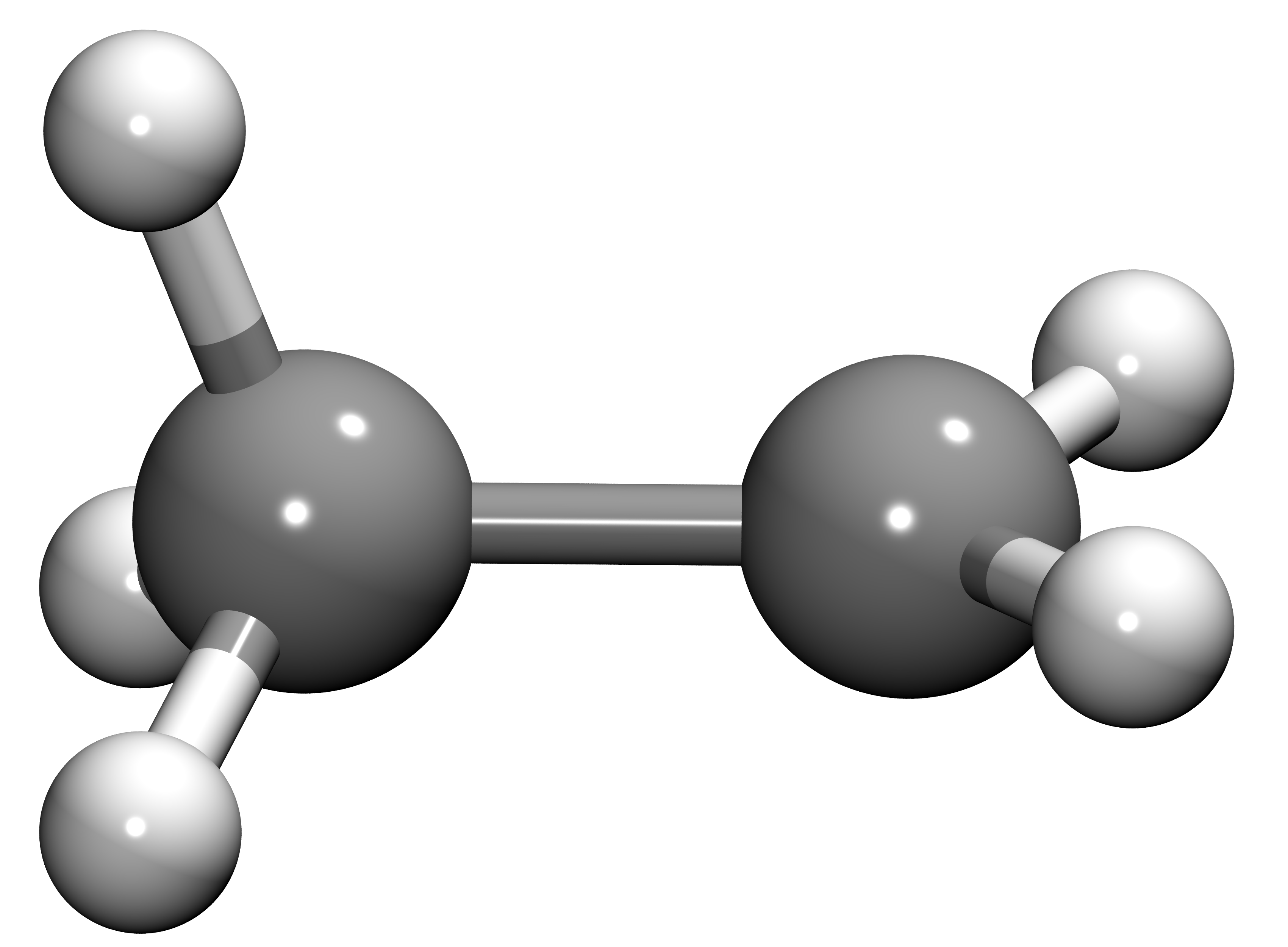}};
\node [right=2em of e0] (cat) {\includegraphics[scale=0.05]{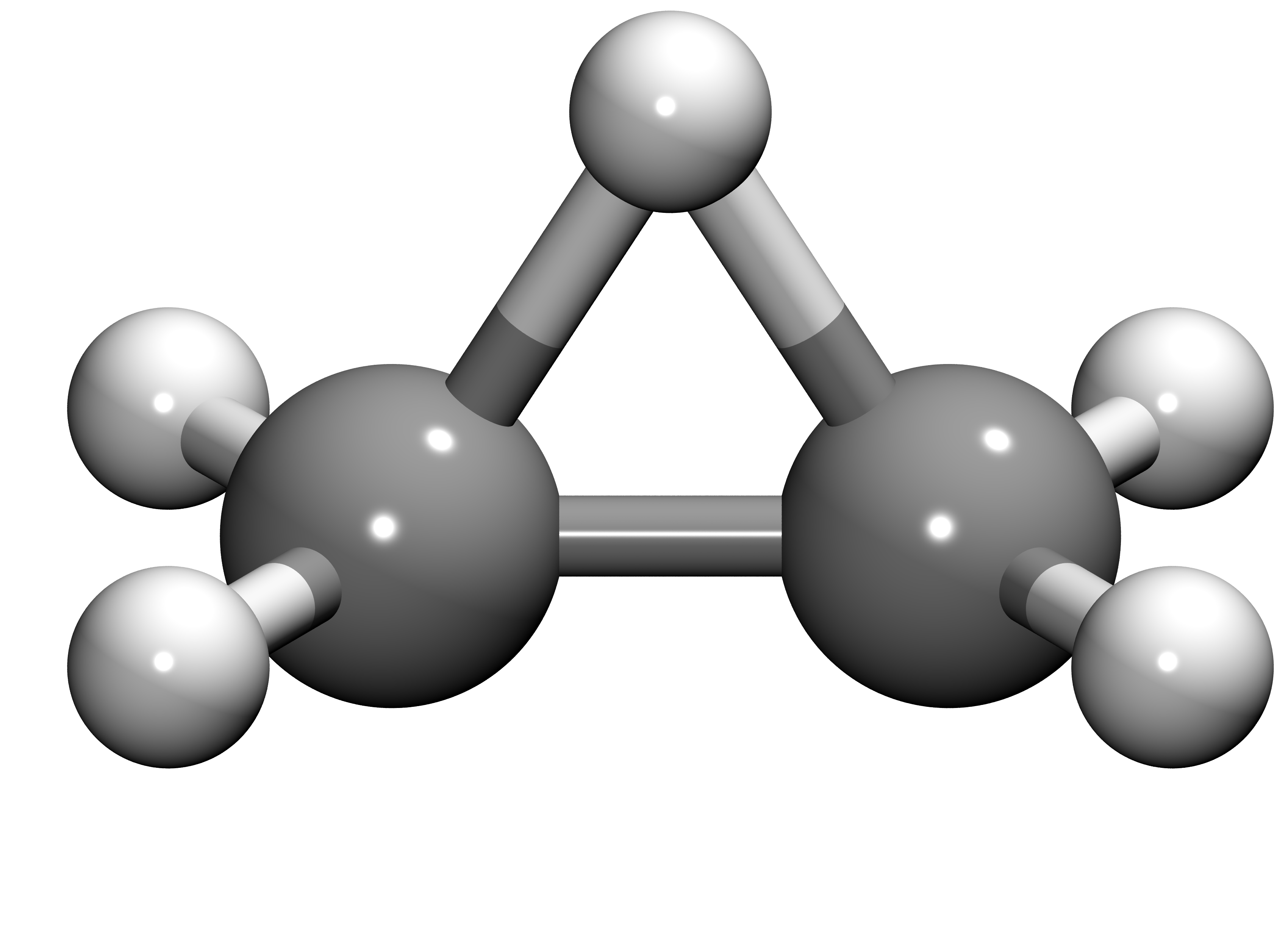}}
    edge[color=white,shorten <=2pt,shorten >=2pt] (net);
\node [below=1.5em of e0] (r0) {(a) $\mathrm{C_2H_5}$ and  $\mathrm{C_2H^+_5}$};
\node [below=5em of e0,fill=white] (e1) {};
\node [left=2em of e1] (net1) {\includegraphics[scale=0.05]{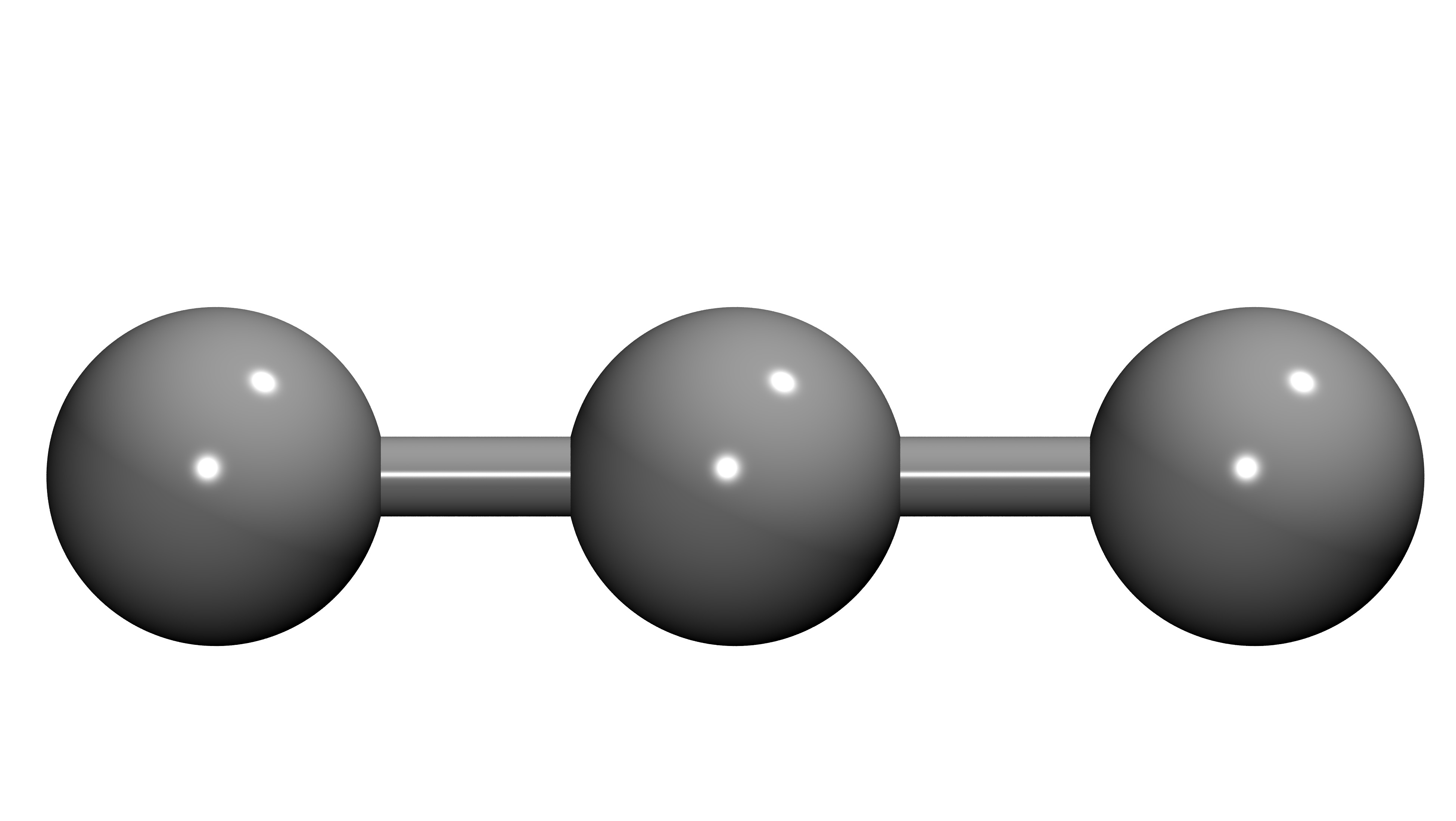}};
\node [right=2em of e1] (cat1) {\includegraphics[scale=0.05]{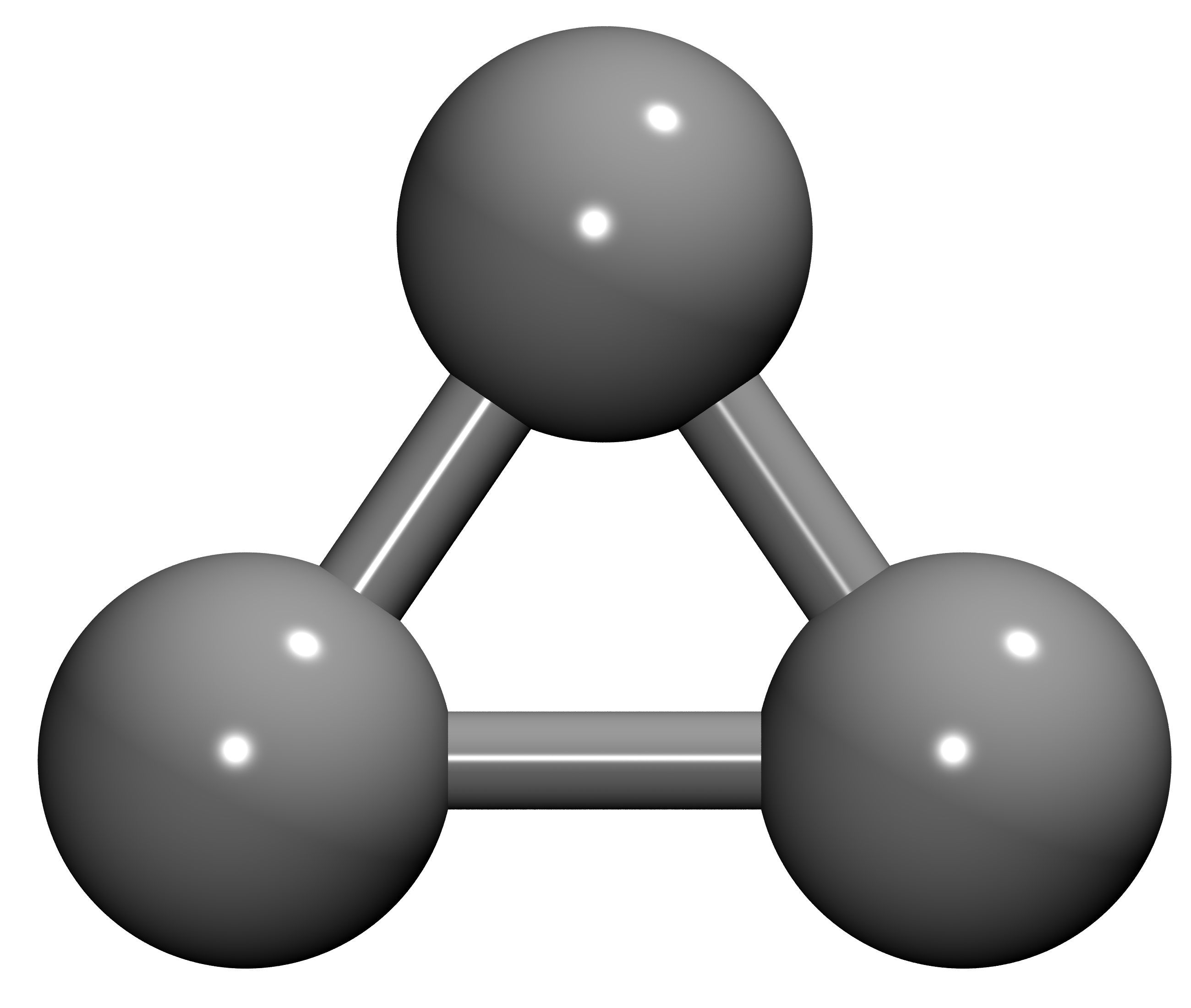}}
    edge[color=white,shorten <=2pt,shorten >=2pt] (net1);
\node [below=1.5em of e1] (r1) {(b) $\mathrm{C_3}$ and $\mathrm{C^+_3}$};
\node [below=7em of e1,fill=white] (e2) {};
\node [left=2em of e2] (net2) {\includegraphics[scale=0.05]{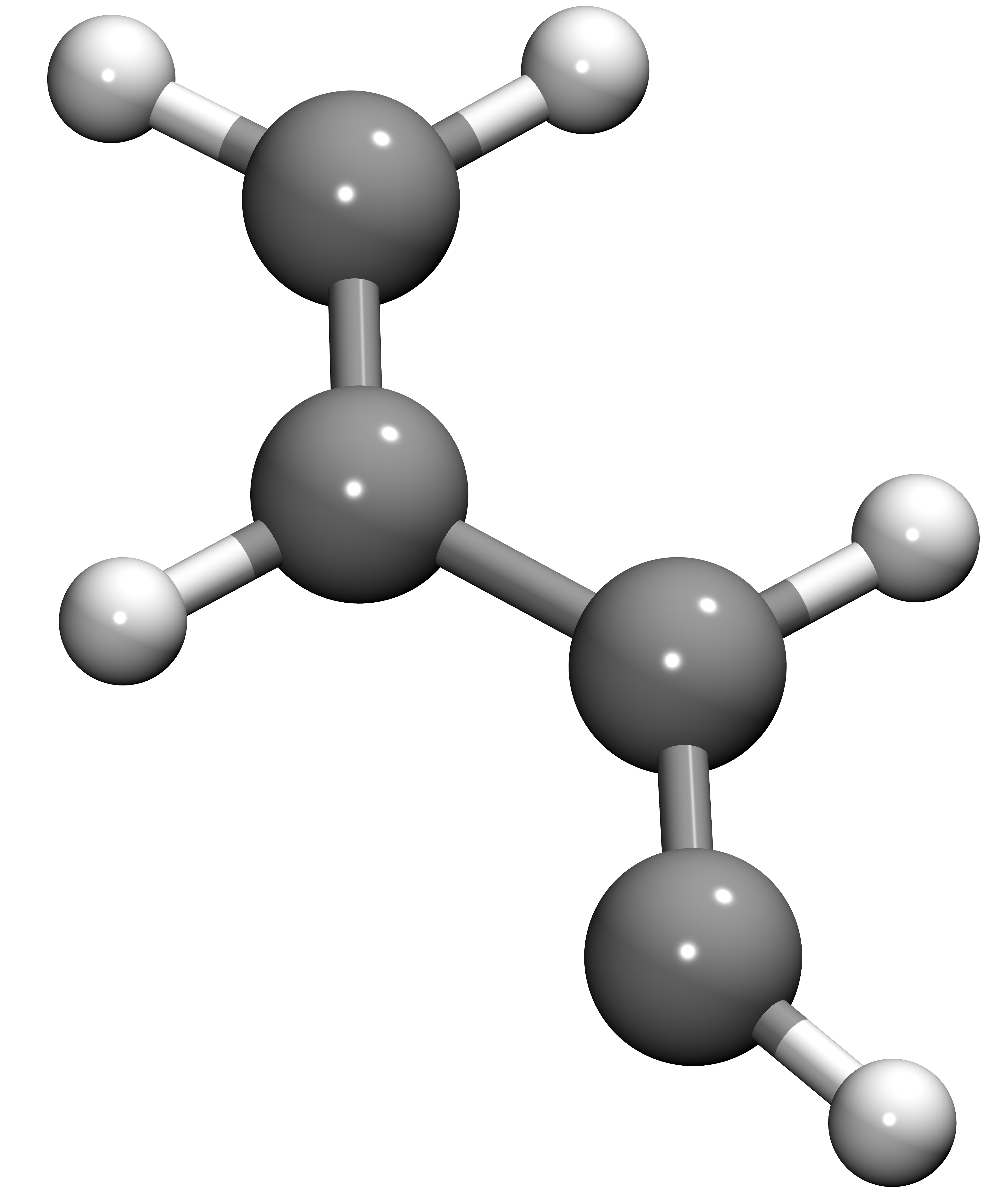}};
\node [right=2em of e2] (cat2) {\includegraphics[scale=0.05]{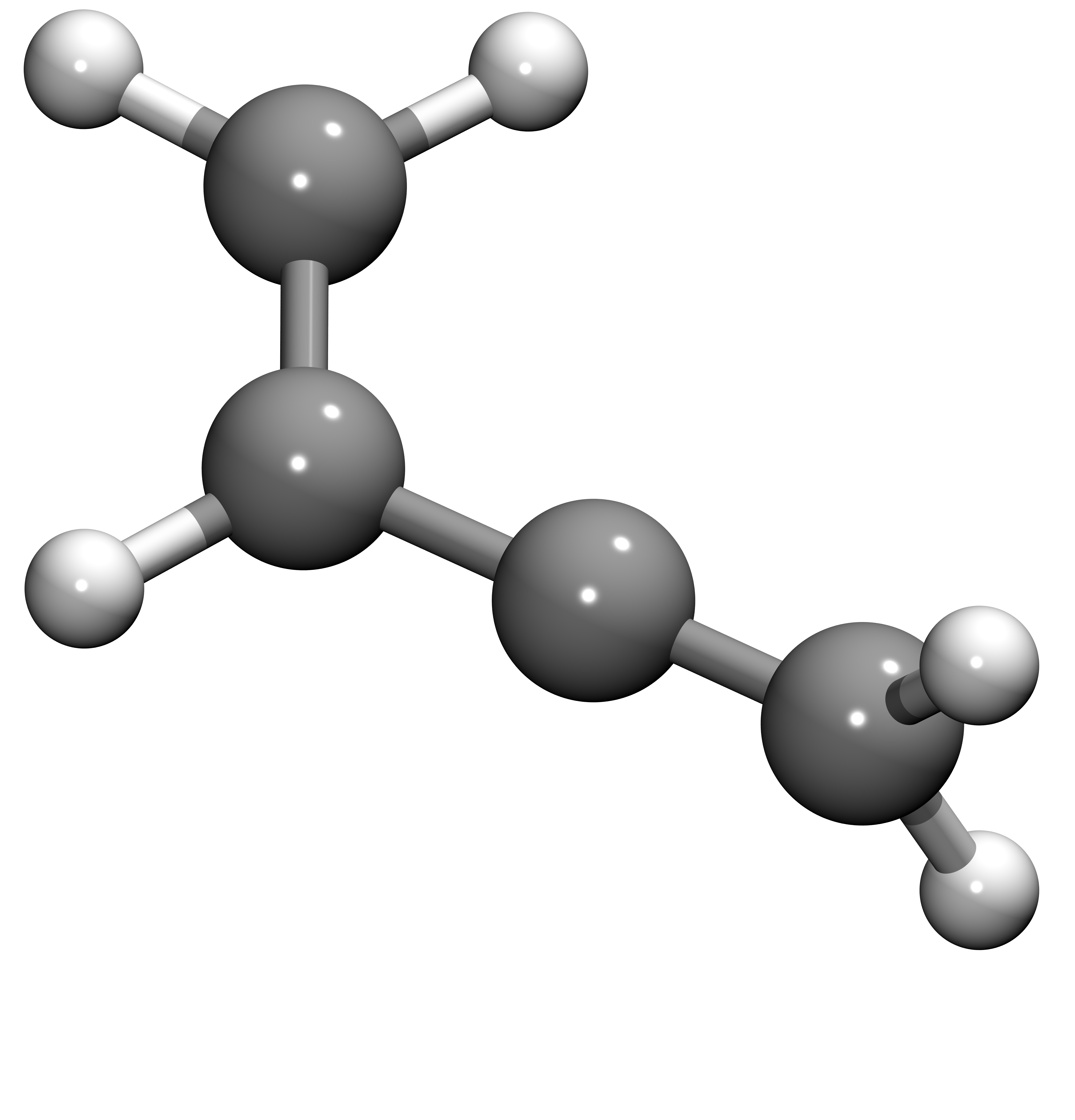}}
    edge[color=white,shorten <=2pt,shorten >=2pt] (net2);
\node [below=3.5em of e2] (r2) {(c) $\mathrm{C_4H_5}$ and $\mathrm{C_4H^+_5}$};
\end{tikzpicture}
\end{center}
\caption{Connectivity change between cations and the neutral species.}  
\label{fig:connectivity}
\end{figure}
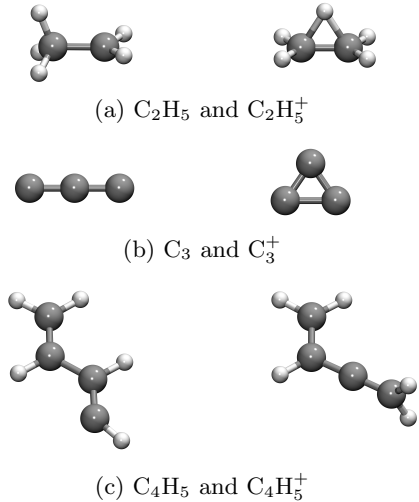

\paragraph{Connectivity change between neutral and cationic partners}
Using the tools described in the subsection \ref{subsec:graph}, the atomic connectivity check was performed on the cations reported in Table \ref{tab:ionization}, and compared to the corresponding neutral species. The only cations that change their connectivity after the ionization and the subsequent geometric optimization are: $\mathrm{C}_3^+$, $\mathrm{C_2H}_5^+$ and $\mathrm{C_4H}_5^+$, as shown in Figure \ref{fig:connectivity}.
For the $\mathrm{C}_3^+$ this rearrangement explains the large difference (${\sim}$1.4 $\mathrm{eV}$) between our computed IE and the experimental vertical IE value. 

This result corroborates our adopted methodology (\S ~\ref{sec:methodology}), because it also demonstrates that the assumption in the astrochemical networks that many cations share the same connectivity of their neutral counterparts is respected.

\section{Discussion}\label{sec:discussion}

As explained in the Introduction, having the physico-chemical properties of the species involved in the astrochemical networks is a first basic step towards achieving accurate and reliable modeling.
In this article, we complemented the work by \cite{woon2009quantum} on neutral species adding the physico-chemical properties of the totality of cations present in the KIDA network$^2$.
In the following, we discuss a first, immediate application of the combined neutral plus cations datasets: how the reaction energy, derived from the neutral and cations datasets, could be used to identify reactions that can not happen in the ISM conditions.

The example regards two reactions, reported in Table \ref{tab:reactions_result}, that involve SiS$^+$.
This cation has been postulated to be a precursor leading to SiS, a species observed in star forming regions and associated with the molecular shocks of young forming protostars \citep{tercero2011,podio2017}. 
Contrarily to the competing S-bearing species SiO, which is likely extracted from the shattered grains, SiS is though to be a product of gas-phase reactions \citep{podio2017}.
The KIDA database only lists one reaction forming SiS: HSiS$^+$ + e $\rightarrow$ SiS + H.
In turn, according to KIDA, HSiS$^+$ is formed by the first reaction in Table \ref{tab:reactions_result}: $\mathrm{H_2 + SiS^+} \rightarrow \mathrm{H + HSiS^+}$.
Our computations, coupled with those by \cite{woon2009quantum}, clearly show that the this reaction is highly endothermic (${\sim}104$ kJ/mol) and, consequently, rule out the formation of SiS by the recombination of HSiS$^+$.
Previous experimental works by \cite{wlodek1989gas} support our conclusion.
Therefore, these two reactions (forming SiS from HSiS$^+$) should be removed by the astrochemical reaction databases.

The second reaction in Table \ref{tab:reactions_result} involves the formation of SiS$^+$, that would be the step before the above SiS reaction formation, according to the KIDA database (on the contrary, the UMIST database does not report the reaction).
Also in this case, our calculations show that the reaction is endothermic (${\sim}36$ kJ/mol) and, therefore, should be removed from the database.

For the curious reader, other routes of SiS formation, involving neutral-neutral reactions, have been explored in the literature since the work of \citep{podio2017} and found to be plausible \citep[][]{rosi2018,zanchet2018,rosi2019electronic}.

\begin{deluxetable}{lc}
\tablehead{Reaction & $\Delta \mathrm{E}$ [kJ/mol]}
\tablecaption{Two example of endothermic reactions found in astrochemical reaction network for molecular cloud. The reaction energy ($i.e.$ $\Delta$E, no ZPE correction) is computed with our and the \cite{woon2009quantum} data.}
\label{tab:reactions_result}
\startdata
$\mathrm{H_2 + SiS^+} \rightarrow \mathrm{H + HSiS^+}$ & 103.6 \\
$\mathrm{SiS + S^+} \; \, \,  \rightarrow \mathrm{SiS^+ + S}$ & 36.1 \\
\enddata
\end{deluxetable}

\section{Conclusions} \label{sec:conclusions}

In this work, we present new \textit{ab-initio} calculations of the structure and energy of 262 cations, all appearing in the used astrochemical reaction network databases (KIDA and UMIST).
Our calculations complement the previous work by \cite{woon2009quantum}, who reported the same properties for an ensemble of 200 neutral species.
The rational behind our new calculations is that the accurate knowledge of the physico-chemical properties of the species in the reaction network databases is a first mandatory step to improve the reliability of the astrochemical models.

All the computed data can be found in the ACO (Astro-Chemistry Origin) project site$^5$ at the link:   \href{https://aco-itn.oapd.inaf.it/aco-public-datasets/theoretical-chemistry-calculations/cations-database}{ACO-Cations-Database}.

Finally, we discussed two practical examples to illustrate the potentiality of using our new cations database, coupled with the \cite{woon2009quantum} one, to identify and exclude endothermic reactions from the astrochemical reaction networks. 

\acknowledgments

This project has received funding within the European Union’s Horizon 2020 research and innovation programme from the European Research Council (ERC) for the project “The Dawn of Organic Chemistry” (DOC), grant agreement No 741002, and from the Marie Sk{\l}odowska-Curie for the project ”Astro-Chemical Origins” (ACO), grant agreement No 811312. 
SP, NB, PU acknowledge the Italian Space Agency for co-funding the Life in Space Project (ASI N. 2019-3-U.O).
CINES-OCCIGEN HPC is kindly acknowledged for the generous allowance of super-computing time through the A0060810797 project. 
Finally, we wish to acknowledge the extremely useful discussions with Prof. Gretobape and the \LaTeX $\,$ community for the insights on TikZ and PGFPlots packages. 

\software{RdKit \citep{landrum2016rdkit}, ASE \citep{larsen2017atomic}, NetworkX \citep{SciPyProceedings_11}, VMD \citep{vmd}, JSmol$^6$, Gaussian16 \citep{g16}, .}

\clearpage

\bibliography{sample63}{}
\bibliographystyle{aasjournal}

\end{document}